\documentclass[twocolumn,showpacs,preprintnumbers,amsmath,amssymb,superscriptaddress,nofootinbib,english]{revtex4-1}
\pdfoutput=1
\usepackage{times,amsmath,amsfonts,amssymb,epstopdf}
\usepackage{graphicx}
\usepackage{dcolumn}
\usepackage{bm}
\usepackage{epsfig}
\usepackage{graphicx}
\usepackage{hyperref}
\usepackage[usenames]{color}
\usepackage{url}
\usepackage[normalem]{ulem}
\usepackage[T1]{fontenc}

\def\be{\begin{equation}}
\def\ee{\end{equation}}
\def\ben{\begin{eqnarray}}
\def\een{\end{eqnarray}}
\def\ba{\begin{array}}
\def\ea{\end{array}}

\newcommand{\bq}{\begin{eqnarray}}
\newcommand{\eq}{\end{eqnarray}}
\newcommand{\bes}{\begin{subequations}}
\newcommand{\ees}{\end{subequations}}

\begin{document}
\newcommand{\half}{{\textstyle\frac{1}{2}}}
\allowdisplaybreaks[3]
\def\triangledown{\nabla}
\def\grad3{\hat{\nabla}}
\def\a{\alpha}
\def\b{\beta}
\def\g{\gamma}\def\G{\Gamma}
\def\d{\delta}\def\D{\Delta}
\def\ep{\epsilon}
\def\et{\eta}
\def\z{\zeta}
\def\t{\theta}\def\T{\Theta}
\def\l{\lambda}\def\L{\Lambda}
\def\m{\mu}
\def\f{\phi}\def\F{\Phi}
\def\n{\nu}
\def\p{\psi}\def\P{\Psi}
\def\r{\rho}
\def\s{\sigma}\def\S{\Sigma}
\def\ta{\tau}
\def\x{\chi}
\def\o{\omega}\def\O{\Omega}
\def\k{\kappa}
\def\pa {\partial}
\def\ov{\over}
\def\br{\\}
\def\ud{\underline}

\newcommand\lsim{\mathrel{\rlap{\lower4pt\hbox{\hskip1pt$\sim$}}
    \raise1pt\hbox{$<$}}}
\newcommand\gsim{\mathrel{\rlap{\lower4pt\hbox{\hskip1pt$\sim$}}
    \raise1pt\hbox{$>$}}}
\newcommand\esim{\mathrel{\rlap{\raise2pt\hbox{\hskip0pt$\sim$}}
    \lower1pt\hbox{$-$}}}
\newcommand{\dpar}[2]{\frac{\partial #1}{\partial #2}}
\newcommand{\sdp}[2]{\frac{\partial ^2 #1}{\partial #2 ^2}}
\newcommand{\dtot}[2]{\frac{d #1}{d #2}}
\newcommand{\sdt}[2]{\frac{d ^2 #1}{d #2 ^2}}    

\title{Halo model and halo properties in Galileon gravity cosmologies}

\author{Alexandre Barreira}
\email[Electronic address: ]{a.m.r.barreira@durham.ac.uk}
\affiliation{Institute for Computational Cosmology, Department of Physics, Durham University, Durham DH1 3LE, U.K.}
\affiliation{Institute for Particle Physics Phenomenology, Department of Physics, Durham University, Durham DH1 3LE, U.K.}

\author{Baojiu Li}
\affiliation{Institute for Computational Cosmology, Department of Physics, Durham University, Durham DH1 3LE, U.K.}

\author{Wojciech A. Hellwing}
\affiliation{Institute for Computational Cosmology, Department of Physics, Durham University, Durham DH1 3LE, U.K.}
\affiliation{Interdisciplinary Centre for Mathematical and Computational Modeling (ICM), University of Warsaw, ul. Pawi\'nskiego 5a, Warsaw, Poland}

\author{Lucas Lombriser}
\affiliation{Institute for Astronomy, University of Edinburgh, Edinburgh, EH9 3HJ, U.K.}

\author{Carlton M. Baugh}
\affiliation{Institute for Computational Cosmology, Department of Physics, Durham University, Durham DH1 3LE, U.K.}

\author{Silvia Pascoli}
\affiliation{Institute for Particle Physics Phenomenology, Department of Physics, Durham University, Durham DH1 3LE, U.K.}

\preprint{IPPP/13/ 103/ DCPT/13/ 208}

\begin{abstract}

We investigate the performance of semi-analytical modelling of large-scale structure in Galileon gravity cosmologies using results from N-body simulations. We focus on the Cubic and Quartic Galileon models that provide a reasonable fit to CMB, SNIa and BAO data. We demonstrate that the Sheth-Tormen mass function and linear halo bias can be calibrated to provide a very good fit to our simulation results. We also find that the halo concentration-mass relation is well fitted by a power law. The nonlinear matter power spectrum computed in the halo model approach is found to be inaccurate in the mildly nonlinear regime, but captures reasonably well the effects of the Vainshtein screening mechanism on small scales. In the Cubic model, the screening mechanism hides essentially all of the effects of the fifth force inside haloes. In the case of the Quartic model, the screening mechanism leaves behind residual modifications to gravity, which make the effective gravitational strength time-varying and smaller than the standard value. Compared to normal gravity, this causes a deficiency of massive haloes and leads to a weaker matter clustering on small scales. For both models, we show that there are realistic halo occupation distributions of Luminous Red Galaxies that can match both the observed large-scale clustering amplitude and the number density of these galaxies.

\end{abstract} 
\maketitle

\section{Introduction}

Over the past few years, models of modified gravity have attracted much attention as an explanation to the observed accelerating expansion of the Universe \cite{Clifton:2011jh}. In these models, the acceleration is a natural consequence of the breakdown of the theory of general relativity (GR) on large scales. This contrasts with models like $\Lambda$CDM, in which GR is the theory of gravity and the acceleration is caused by a mysterious "dark energy" component, such as a cosmological constant $\Lambda$ or a slowly-rolling, minimally coupled scalar field. Although these two classes of models can be set up to have identical expansion histories, they will typically differ in how matter and light react to the distribution of gravitational sources. Hence, observables associated with the growth rate of structure and lensing have the potential to test the law of gravity on cosmological scales.

Here, we focus on Galileon gravity models, which were first proposed in Ref.~\cite{PhysRevD.79.064036}. In particular, Ref.~\cite{PhysRevD.79.064036} showed that there are only five Lagrangian density terms ($\mathcal{L}_i, i = 1,...,5$) for a single scalar field $\varphi$ (i) whose physics is invariant under the so-called Galilean shift transformation $\partial_\mu\varphi \rightarrow \partial_\mu\varphi + b_\mu$ (where $b_\mu$ is a constant four-vector); and (ii) whose equations of motion in flat spacetime are up to second order in field derivatives. The Lagrangian terms are characterized by the power with which $\varphi$ (from hereon in the {\it Galileon field}) appears. The linear ($\mathcal{L}_1$) and quadratic ($\mathcal{L}_2$) terms correspond to a linear potential function and to the canonical kinetic term, respectively. The remaining cubic ($\mathcal{L}_3$), quartic ($\mathcal{L}_4$) and quintic ($\mathcal{L}_5$) terms contain nonlinear derivative self-couplings of the Galileon field, which are behind the modifications to gravity. In a Friedmann-Robertson-Walker (FRW) spacetime, however, the original action of Ref.~\cite{PhysRevD.79.064036} leads to higher-order equations of motion, and therefore, to the propagation of Ostrogradski ghosts \cite{Woodard:2006nt}. This problem was fixed in Refs.~\cite{PhysRevD.79.084003, Deffayet:2009mn} by introducing explicit couplings between the Galileon derivative terms and curvature tensors in $\mathcal{L}_4$ and $\mathcal{L}_5$. This, however, breaks the Galilean shift symmetry. The Galileon model is therefore the sector of the general Horndeski theory \cite{Horndeski:1974wa} that is Galilean invariant in the limit of flat spacetime.

A vital requirement of modified gravity models is that any departure from GR on small scales has to satisfy the stringent Solar System tests of gravity \cite{Will:2005va}. This is usually realized by invoking a screening mechanism that dynamically suppresses the modifications to gravity in regions where the density is high, like the Solar System. Interestingly, the nonlinear nature of the derivative couplings in $\mathcal{L}_{3-5}$ that drive the modifications to gravity is also what allows these modifications to be suppressed in high density regions. This is a mechanism that is widely known as the Vainshtein effect \cite{Vainshtein1972393, Babichev:2013usa, Koyama:2013paa}. The general picture is as follows. In regions where the overdensity is small, the nonlinearities are negligible and the perturbed Galileon field equation of motion essentially becomes a linear Poisson-like equation. Therefore, the Galileon field perturbation acts like an extra gravitational potential and its spatial gradient gives rise to a sizeable fifth force. On the other hand, closer to massive bodies, as the density gets higher, the nonlinear terms become increasingly important, effectively suppressing the magnitude of the fifth force so that normal gravity is felt.

Many recent papers have studied the cosmological properties of the covariant Galileon model at the linear level in perturbation theory \cite{Barreira:2012kk, Barreira:2013jma, Gannouji:2010au, PhysRevD.80.024037, DeFelice:2010pv, Nesseris:2010pc, Appleby:2011aa, PhysRevD.82.103015, Neveu:2013mfa, Appleby:2012ba, Okada:2012mn, Bartolo:2013ws}. In particular, the work of Refs.~\cite{Barreira:2012kk, Barreira:2013jma} has shown that the Galileon model can fit the Cosmic Microwave Background (CMB) data better than $\Lambda$CDM, because it allows for less power on the largest angular scales, which is slightly favoured by the measurements from the WMAP \cite{Hinshaw:2012fq} and Planck \cite{Ade:2013zuv} satellites. {However, Refs.~\cite{Barreira:2012kk, Barreira:2013jma} have also discovered that the predicted amplitude of the linear matter power spectum of the Galileon models that best fit the CMB data is substantially larger than that in $\Lambda$CDM. This raised the possibility that the Galileon predictions may be in tension with the observed large-scale galaxy distribution \cite{Reid:2009xm}, which we investigate here.}

The above papers considered only linear perturbation theory which, by definition, ignores the effects of the intrinsically nonlinear Vainshtein screening. Moreover, a realistic comparison between theory and observations is also subject to a proper understanding of how the distribution of galaxies and their host haloes is biased relative to the underlying dark matter density field. To address these uncertainties, one needs to study the nonlinear evolution of the matter and Galileon field perturbations. The most accurate way to do this is through N-body simulation. References~\cite{Barreira:2013eea} and \cite{Li:2013tda} performed the first N-body simulations of the covariant Galileon model of Ref.~\cite{PhysRevD.79.084003}. These two works focused on the so-called Cubic ($\mathcal{L}_{2-3}$) and Quartic ($\mathcal{L}_{2-4}$) Galileon models (see Sec.~\ref{sec:model}), respectively. They found that the nonlinearities of the Vainshtein screening, although very noticeable on small scales, do not have an impact on the clustering of matter on scales $k \lesssim 0.1 h/{\rm Mpc}$. N-body simulations of the most general Quintic model ($\mathcal{L}_{2-5}$) are more challenging to perform. Nevertheless, by studying the behavior of the fifth force assuming spherical symmetry {in the quasi-static limit}, Ref.~\cite{Barreira:2013xea} has shown that the equations of motion of the Quintic model fail to provide physical solutions when the density perturbations become of order unity. Consequently, the study of nonlinear structure formation in the Quintic model is no longer of interest. By making use of the spherical collapse model and the excursion set theory formalism, Ref.~\cite{Barreira:2013xea} also estimated the predicted halo mass function and linear halo bias in the Quartic model. The indications were that the halo bias could be smaller than in standard $\Lambda$CDM models. However, because of the simplified treatment, no decisive quantitative statements could be made.

Here, our goal is to put the excursion set theory predictions of Ref.~\cite{Barreira:2013xea} on to a more quantitative level. In particular, by comparing to the results of N-body simulations, we analyse the performance of the fitting formulae based on the ellipsoidal collapse of structures to describe the halo mass function and linear bias \cite {Sheth:1999mn, Sheth:1999su, Sheth:2001dp}. We anticipate that these fitting formulae can match the simulation results very well, provided their adjustable parameters are calibrated against the simulations. We also measure and fit the halo concentration parameter. {Ultimately,} we use these halo properties to determine the nonlinear matter power spectrum computed in the halo model approach \cite{Cooray:2002dia}. The study of these simplified analytical formulae helps us to understand better the physical picture of the halo distribution, which is often hidden in the brute-force calculations of a N-body simulation. These formulae also allow for an efficient exploration of the models parameter space, which would not be possible with the time-consuming N-body simulations.

This paper is organized as follows. In Sec.~\ref{sec:model}, we display the action, the cosmological parameters and the equations that govern the gravitational interaction in spherically symmetric configurations of the Galileon models we consider. In Sec.~\ref{sec:halo-model-formulae}, we present the equations of the halo model of the nonlinear matter power spectrum. We also describe the Sheth-Tormen formulae for the halo mass function and linear halo bias \cite {Sheth:1999mn, Sheth:1999su, Sheth:2001dp}, and the Navarro-Frenk-White (NFW) \cite{Navarro:1996gj} concentration parameter of dark matter haloes. In Sec.~\ref{sec:results}, we show our main results, where we compare {and calibrate} the analytical predictions with the results of the N-body simulations. {We also use the calibrated formulae for the halo mass function and linear bias to conduct a halo occupation distribution analysis of Luminous Red Galaxies (LRGs).} Finally, we summarize and draw our conclusions in Sec.~\ref{sec:conclusions}.

Unless otherwise specified, we assume the metric convention $(+,-,-,-)$ and work in units in which the speed of light $c = 1$. Greek indices run over $0,1,2,3$ and we use $8\pi G=\kappa=M^{-2}_{\rm Pl}$ interchangeably, where $G$ is Newton's constant and $M_{\rm Pl}$ is the reduced Planck mass.

\section{The Galileon model}\label{sec:model}

\subsection{Action}

The action of the covariant Galileon model \cite {PhysRevD.79.084003} is given by

\bq\label{eq:action}
&& S = \int {\rm d}^4x\sqrt{-g} \left[ \frac{R}{16\pi G} - \frac{1}{2}\sum_{i=1}^5c_i\mathcal{L}_i - \mathcal{L}_m\right],
\eq
where $R$ is the Ricci curvature scalar and $g$ is the determinant of the metric $g_{\mu\nu}$. $\mathcal{L}_m$ describes the matter content of the universe, which in this model is minimally coupled to the metric and Galileon fields. The parameters $c_{1-5}$ are real dimensionless constants and the five Lagrangian density terms, fixed by the Galilean invariance in flat spacetime, $\partial_\mu\varphi \rightarrow \partial_\mu\varphi + b_\mu$, are given by
\bq\label{L's}
\mathcal{L}_1 &=& M^3\varphi, \nonumber \\
\mathcal{L}_2 &=& \nabla_\lambda\varphi\nabla^\lambda\varphi,  \nonumber \\
\mathcal{L}_3 &=& \frac{2}{M^3}\Box\varphi\nabla_\lambda\varphi\nabla^\lambda\varphi, \nonumber \\
\mathcal{L}_4 &=& \frac{1}{M^6}\nabla_\lambda\varphi\nabla^\lambda\varphi\Big[ 2(\Box\varphi)^2 - 2(\nabla_\mu\nabla_\nu\varphi)(\nabla^\mu\nabla^\nu\varphi) \nonumber \\
&& -R\nabla_\mu\varphi\nabla^\mu\varphi/2\Big], \nonumber \\
\mathcal{L}_5 &=&  \frac{1}{M^9}\nabla_\lambda\varphi\nabla^\lambda\varphi\Big[ (\Box\varphi)^3 - 3(\Box\varphi)(\nabla_\mu\nabla_\nu\varphi)(\nabla^\mu\nabla^\nu\varphi) \nonumber \\
&& + 2(\nabla_\mu\nabla^\nu\varphi)(\nabla_\nu\nabla^\rho\varphi)(\nabla_\rho\nabla^\mu\varphi) \nonumber \\
&& -6 (\nabla_\mu\varphi)(\nabla^\mu\nabla^\nu\varphi)(\nabla^\rho\varphi)G_{\nu\rho}\Big],
\eq
in which $M^3\equiv M_{\rm Pl}H_0^2$, where $H_0$ is the Hubble expansion rate today. In this model, the nonlinear coupling of the covariant derivatives of $\varphi$ induces interactions between partial derivatives of $g_{\mu\nu}$ and $\varphi$ (a process known as kinetic gravity braiding \cite{Deffayet:2010qz, Babichev:2012re, Kimura:2010di}). This is why this model is a modified gravity model. In addition to these interactions, there are also direct couplings to the Ricci scalar, $R$, and the Einstein tensor, $G_{\mu\nu}$, in $\mathcal{L}_4$ and $\mathcal{L}_5$, respectively. These latter two couplings are needed so that the equations of motion are kept up to second-order in field derivatives in curved spacetimes, such as the one described by the FRW metric \cite{PhysRevD.79.084003}. However, this process comes at the cost of breaking the Galilean shift symmetry.

We will set the potential term $\mathcal{L}_1$ to zero (i.e. $c_1 = 0$) as we are only interested in cases where the acceleration is driven solely by kinetic energy terms. Note that in this case, the value of $\varphi$ becomes irrelevant for the physics of the model since the action contains only $\partial\varphi$ terms. One can then consider three classes of Galileon models. The Cubic Galileon model is the simplest and encompasses $\mathcal{L}_2$ and $\mathcal{L}_3$ only; the Quartic Galileon model considers $\mathcal{L}_{2-4}$; and finally, the Quintic Galileon is the most general which includes all of the Lagrangian terms $\mathcal{L}_{2-5}$. It is worth noting that none of the Cubic, Quartic or Quintic models possess a $\Lambda$CDM limit for the expansion history of the Universe. Recently, Ref.~\cite{Barreira:2013xea} has shown that, if the density perturbations become higher than $\mathcal{O}(1)$, then the equation of motion of the scalar field in the Quintic models that provide a good fit to the CMB data fails to admit real physical solutions for the spatial gradient of $\varphi$ in spherically symmetric {quasi-static} configurations. Naturally, this prevents one from studying nonlinear structure formation in the Quintic model. As a result, in what follows, we always assume that $c_5 = 0$ and focus only on the Cubic and Quartic Galileon models.

By varying the action of Eq.~(\ref{eq:action}) with respect to $g_{\mu\nu}$ and $\varphi$, one obtains the modified Einstein equations and the Galileon field equation of motion, respectively. These equations are lengthy and therefore we do not show them in this paper (the interested reader can find them in Eqs.~(A1-A7) of Ref.~\cite{Barreira:2012kk}). The background and the linearly perturbed equations have also been derived and presented in previous papers (e.g.~Refs.~\cite{Barreira:2012kk, Barreira:2013jma}), and we also abstain from showing them here. In this paper, we simply layout the nonlinear equations that determine the modifications to gravity in spherical matter overdensities.

\subsection{Fifth force solutions}

\begin{table}
\caption{Parameters of the Galileon models studied in this paper. $\Omega_{r0}$, $\Omega_{b0}$, $\Omega_{c0}$, $h$, $n_s$, and $\tau$ are, respectively, the present-day fractional energy density of radiation ($r$), baryons ($b$) and cold dark matter ($c$), the dimensionless present-day Hubble expansion rate, the primordial scalar spectral index and the optical depth to reionization. The scalar amplitude at recombination $A_s$ refers to a pivot scale $k = 0.02 {\rm Mpc}^{-1}$. The universe is spatially flat in these models. The parameters $c_2$, $c_3$, $c_4$, $c_5$ are the dimensionless constants that appear in the action Eq.~(\ref{eq:action}) and $\rho_{\varphi, i} / \rho_{m,i}$ is the ratio of the Galileon and total matter ($m$) energy densities at $z_i$. We also show the Galileon field time derivative $\dot{\bar{\varphi}}_i c_3^{1/3}$ at $z_i$ , the age of the Universe and the present-day value of $\sigma_8$. Only in this table, the subscript "$_i$" refers to quantities evaluated at $z = z_i = 10^6$. {The predicted CMB angular power spectrum and the linear matter power spectrum of these model parameters can be found in Fig.~1 of Ref.~\cite{Barreira:2013xea}.}}
\begin{tabular}{@{}lccccccccccc}
\hline\hline
\\
Parameter  & Cubic Galileon& \ \ Quartic Galileon& \ \ 
\\
\hline
\\
$\Omega_{r0}{h}^2$                  &     $4.28\times10^{-5}$          				&\ \ $4.28\times10^{-5}$ & \ \ 
\\
$\Omega_{b0}{h}^2$                     & $0.02196$           				&\ \ $0.02182$ & \ \ 
\\
$\Omega_{c0}{h}^2$                       & $0.127$          				&\ \   $0.126$ & \ \ 
\\
${h}$                                                    & $0.731$            			&\ \   $0.733$ & \ \ 
\\
$n_s$                                                   & $0.953$             			           &\ \    $0.945$ & \ \ 
\\
$\tau$                                                   & $0.0763$                  			&\ \   $0.0791$ & \ \
\\
${\rm log}\left[ 10^{10}A_s \right]$    & $3.154$			&\ \  $3.152$ & \ \ 
\\
${\rm log}\left[\rho_{\varphi, i} / \rho_{m,i}\right]$   &$-4.22$      &\ \   $-37.39$ & \ \ 
\\
$c_2 / c_3^{2/3}$                                                    &$-5.38$                               &\ \    $-4.55$ & \ \
\\
$c_3$                                                                        &$10$           &\ \    $20$ & \ \
\\
$c_4/ c_3^{4/3}$                                                  &$0$ (fixed)                                 &\ \    $-0.096$ & \ \
\\
$c_5/ c_3^{5/3}$                                                  &$0$ (fixed)                                 &\ \    $0$ (fixed) & \ \
\\
\hline
\\
$\dot{\bar{\varphi}}_i c_3^{1/3}$		&$1.10\times10^{-9}$			  		&\ \  $1.54\times10^{-20}$ & \ \
\\
Age (Gyr)					  	&$ 13.748$	&\ \ $13.770$ & \ \
\\
$\sigma_8(z = 0)$					  &$0.997$		&\ \ $0.998$ & \ \
\\
\hline
\hline
\end{tabular}
\label{table:table-max}
\end{table}

\begin{figure*}
	\centering
	\includegraphics[scale=0.455]{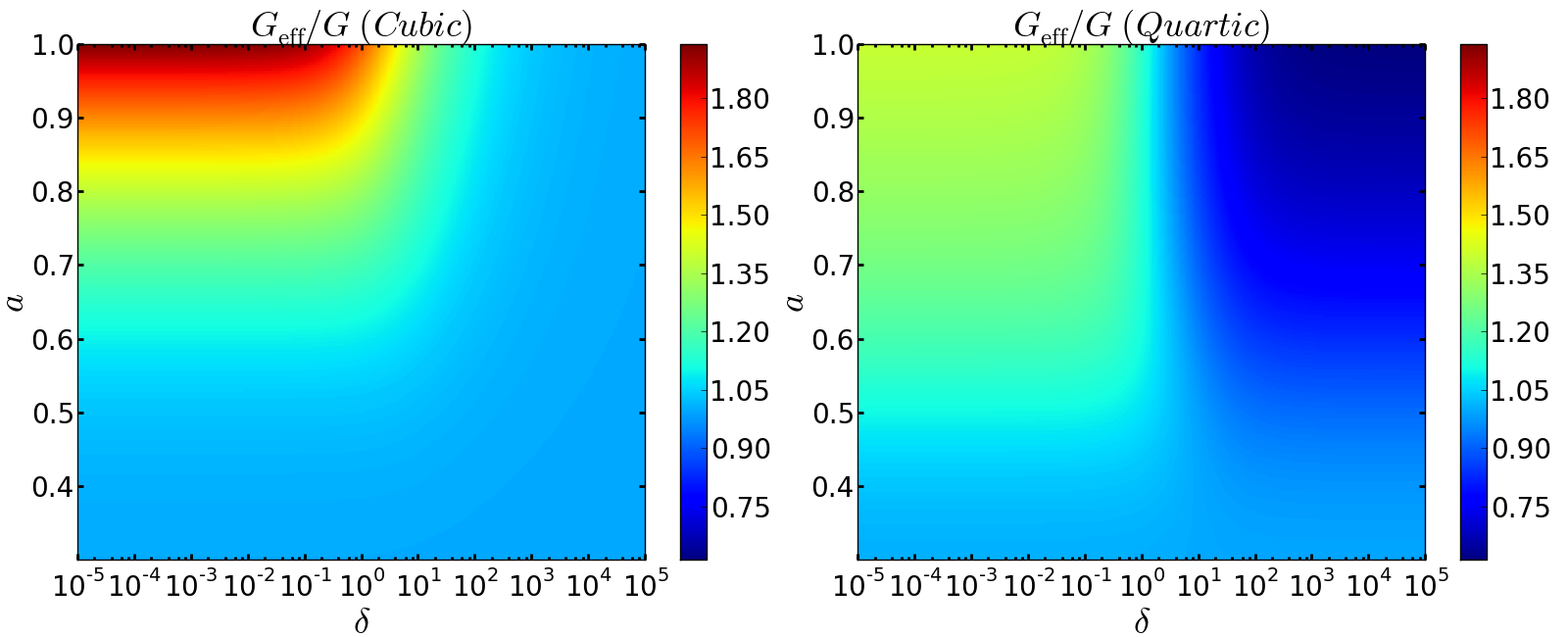}
	\caption{Time and density dependence of the effective gravitational strength $G_{\rm eff}$ of Eq.~(\ref{eq:Geff}) for the Cubic (left panel) and Quartic (right panel) Galileon models. The colour scale bars at the right of each panel show the value of $G_{\rm eff}/G$. The color scale is the same for both panels.}
\label{fig:Geffmap}\end{figure*}

\begin{table}
\caption{Values of the critical initial overdensity for the spherical collapse to occur at $a = 0.6$, $a = 0.8$, $a = 1.0$, extrapolated to $a = 1.0$ with the $\Lambda$CDM linear growth factor. This extrapolation is done purely to enable the resulting values of $\delta_c$ in the Galileon model to be more easily compared to results from other models in the literature. The values of $\delta_c$ are shown for $\Lambda$CDM and the models of Table \ref{table-variants}, and were obtained by following the strategy presented in Ref.~\cite{Barreira:2013xea}.}
\begin{tabular}{@{}lccccccccccc}
\hline\hline
\\
Model  & \ \ $a = 0.6$ & $a = 0.8$ & $a = 1.0$ &\ \ 
\\
               & \ \ $\delta_c$ & $\delta_c$ & $\delta_c$ &\ \ 
\\
\hline
\\
$\Lambda$CDM                                         &\ \  $2.346$ &  \ \ $1.907$ & $1.677$&\ \ 
\\
\\
${\rm QCDM}_{\rm Cubic}$                    &\ \  $2.224$ &  \ \ $1.780$ & $1.560$&\ \ 
\\
Cubic Galileon                                             &\ \  $2.219$ &  \ \ $1.767$ & $1.537$&\ \ 
\\
Linearized Cubic Galileon                        &\ \  $2.213$ &  \ \ $1.749$ & $1.500$&\ \ 
\\
\\
${\rm QCDM}_{\rm Quartic}$                 &\ \  $2.226$ &  \ \ $1.784$ & $1.565$&\ \ 
\\
Quartic Galileon                                          &\ \  $2.236$ &  \ \ $1.805$ & $1.593$&\ \ 
\\
Linearized Quartic Galileon                    &\ \  $2.196$ &  \ \ $1.728$ & $1.492$&\ \ 
\\
\hline
\hline
\end{tabular}
\label{table:dc-values}
\end{table}

To build some intuition about the results that are presented in the subsequent sections, it is instructive to look at the modifications to gravity in the limit of spherical symmetry, for which analytical solutions can be derived. In this case, according to Ref.~\cite{Barreira:2013xea}, the total gravitational force in the Galileon model is obtained by solving the following equations

\bq
\label{eq:poisson}\frac{\Phi,_{\chi}}{\chi} &=& \frac{\Omega_{m0}\delta a^{-3} + A_1\left(\varphi,_{\chi}/\chi\right) + A_2\left(\varphi_{\chi}/\chi\right)^2}{A_4}, \\
\label{eq:slip}\frac{\Psi,_{\chi}}{\chi} &=& \frac{B_0\left(\Phi,_{\chi}/\chi\right) + B_1\left(\varphi,_{\chi}/\chi\right) + B_2\left(\varphi,_{\chi}/\chi\right)^2}{B_3}, \\
\label{eq:eom_sph_alg} 0 &=& \eta_{01}\delta + \left(\eta_{11}\delta + \eta_{10}\right)\left[\frac{\varphi,_{\chi}}{\chi}\right] +\eta_{20}\left[\frac{\varphi,_{\chi}}{\chi}\right]^2 \nonumber \\
&& + \eta_{30}\left[\frac{\varphi,_{\chi}}{\chi}\right]^3,
\eq
where $\chi \equiv aH_0^2r$ (with $r$ being the comoving radial coordinate and $a$ the cosmic scale factor) and $\delta \equiv \rho_m/\bar{\rho}_m - 1$ is the matter density contrast of a top-hat spherical overdensity w.r.t. the mean background density $\bar{\rho}_m$. The quantities $A_i$, $B_i$ and $\eta_{ab}$ depend only on time and can be found in Ref.~\cite{Barreira:2013xea}. The two gravitational potentials $\Phi$ and $\Psi$ are defined by the perturbed spatially-flat FRW line element in the Newtonian gauge

\bq\label{metric}
{\rm d}s^2 = \left(1 + 2\Psi\right){\rm d}t^2 - a(t)^2\left(1 - 2\Phi\right)\gamma_{ij}{\rm d}x^i{\rm d}x^j,
\eq
where $\gamma_{ij} = \rm{diag}\left[1, 1, 1\right]$. Step by step, for fixed $a$ and $\delta$, one can solve Eq.~(\ref{eq:eom_sph_alg}) analytically for ${\varphi,_{\chi}}/{\chi}$, and plug the solution into Eqs.~(\ref{eq:poisson}) and (\ref{eq:slip}) to determine the total force given by ${\Psi,_{\chi}}/{\chi}$. Note that we have written Eqs.~(\ref{eq:poisson}), (\ref{eq:slip}) and (\ref{eq:eom_sph_alg}) assuming $c_5 = 0$. If $c_5 \neq 0$, Eq.~(\ref{eq:eom_sph_alg}) becomes a sixth-order algebraic equation for ${\varphi,_{\chi}}/{\chi}$, which does not admit an analytical solution and has to be solved numerically. Moreover, in the case of the Cubic Galileon, the equations simplify even further since $\eta_{11}, \eta_{30}, A_2, B_1, B_2$ vanish and $B_0 = B_3 = -2$. In particular, in the Cubic model $\Phi = \Psi$.

Table \ref{table:table-max} lists the model parameters we consider. These were obtained by following the steps in Ref.~\cite{Barreira:2013jma}, which used data from the WMAP 9-yr results for the temperature fluctuations power spectrum of the CMB \cite{Hinshaw:2012fq}, type Ia supernovae (SNIa) from the SNLS 3-yr sample \cite{Guy:2010bc} and baryonic acoustic oscillations (BAO) measurements from the 6dF \cite{Beutler:2011hx}, SDSS DR7 \cite{Reid:2012sw} and BOSS \cite{Percival:2009xn} galaxy surveys. These correspond also to the parameters used to run the N-body simulations in Refs.~\cite{Barreira:2013eea} and \cite{Li:2013tda}. The predicted CMB angular power spectrum and the linear matter power spectrum of these model parameters can be found in Fig.~1 of Ref.~\cite{Barreira:2013xea}. In the rest of the paper, whenever we refer to the Cubic and Quartic Galileon models we mean these particular parameter sets.

The modifications to gravity can be quantified in terms of an effective gravitational strength $G_{\rm eff}$ defined as

\bq\label{eq:Geff}
\frac{G_{\rm{eff}}}{G}(a, \delta) = \frac{\Psi,_{\chi}/\chi}{\Psi,_{\chi}^{\rm GR}/\chi} = \frac{\Psi,_{\chi}/\chi}{\Omega_{m0}\delta /\left(2a^3\right)}.
\eq
In Ref.~\cite{Barreira:2013xea}, this quantity is shown for the Quartic model as a function of time and density. In Fig.~\ref{fig:Geffmap}, for completeness, we repeat the same figure but include also the corresponding result for the Cubic Galileon. In both models, the deviation of $G_{\rm eff}/G$ from unity arises only at late times: $a \gtrsim 0.6$ and $a \gtrsim 0.5$ for the Cubic and Quartic models, respectively. In the linear regime ($|\delta| \ll 1$), $G_{\rm eff}$ increases with time in both models, but it does so more pronouncedly in the Cubic Galileon: at $a = 1$, $G_{\rm eff}$ is roughly $90\%$ larger than $G$ in the Cubic, but only $40\%$ larger in the Quartic. In higher-density regions ($\delta \gtrsim 1$), however, the qualitative pictures of these two models become distinct. Focusing for instance on the present day ($a = 1$), in the Cubic model, $G_{\rm eff}/G$ approaches unity as the density contrast increases. This shows the effect of the screening mechanism in high-density regions. On the other hand, in the Quartic model, at $a = 1$, $G_{\rm eff}$ does not approach the standard value, but instead becomes roughly $40\%$ smaller and time-varying (the time variation can also be seen in Fig.~4 of Ref.~\cite{Li:2013tda}). This result puts the model into severe tension with Solar System tests of gravity that constrain the time variation of $G_{\rm eff}/G$ to be very small \cite{Will:2005va}. The weaker gravity in the Quartic Galileon model follows from modifications induced by the time-varying $A_4$, $B_0$ and $B_3$ terms in Eqs.~(\ref{eq:poisson}) and (\ref{eq:slip}). These terms do not depend on the spatial gradients of the Galileon field, and therefore, cannot be suppressed by the Vainshtein mechanism. The origin of these three coefficients can be traced back to the direct coupling of the Galileon field with the Ricci scalar $R$ in $\mathcal{L}_4$, which is necessary to avoid the presence of ghosts.  In the case of the Cubic Galileon, these three terms are constant and the problem does not arise.

The physical picture depicted in Fig.~\ref{fig:Geffmap} indicates that the modelling of halo properties can be very different in these two models, in particular because of the different behavior in high-density regions. In the remainder of the paper, we focus on these differences, ignoring for now the fact that the time-varying $G_{\rm eff}$ in high-density regions is putting the Quartic model into huge observational tension.

\section{Halo Model of the nonlinear matter power spectrum}\label{sec:halo-model-formulae}

In this section, we describe the halo model of the nonlinear matter power spectrum, as well as the halo properties that are needed as input. In particular, we define and present the halo mass function, linear halo bias and halo density profiles.

\subsection{Halo model}

In the halo model approach, one of the main premises is that all matter in the Universe is in bound structures. Thus, the two-point correlation function of the matter density field can be decomposed into the contributions from the correlations between mass elements that belong to the same halo (the 1-halo term) and to different haloes (the 2-halo term). In terms of the matter power spectrum, this can be written as (see Ref.~\cite{Cooray:2002dia} for a comprehensive review)

\bq\label{eq:halo-model}
P_k = P_k^{\rm 1h} + P_k^{\rm 2h},
\eq
where

\bq\label{eq:halo-model-terms}
P_k^{\rm 1h} &=& \int {\rm d}M \frac{M}{\bar{\rho}_{m0}^2}\frac{{\rm d}n(M)}{{\rm dln}M} |u(k, M)|^2, \nonumber \\
P_k^{\rm 2h} &=& I(k)^2P_{k,{\rm lin}} ,
\eq
are, respectively, the 1-halo and 2-halo terms, with 

\bq\label{eq:I-function}
I(k) = \int {\rm d}M \frac{1}{\bar{\rho}_{m0}}\frac{{\rm d}n(M)}{{\rm dln}M} b_{\rm lin}(M) |u(k, M)|.
\eq

In the above expressions, $k$ is the comoving wavenumber; $\bar{\rho}_{m0}$ is the present-day background matter density; $P_{k, \rm lin}$ is the matter power spectrum obtained using linear theory; ${{\rm d}n(M)}/{{\rm dln}M}$ denotes the comoving number density of haloes per differential logarithmic interval of mass (we shall refer to this quantity as the {\it mass function}); $b_{\rm lin}(M)$ is the linear halo bias; $u(k, M)$ is the Fourier transform of the density profile of the haloes truncated at the size of the halo and normalized such that $u(k \rightarrow 0, M) \rightarrow 1$. In order to compute the matter power spectrum of Eq.~(\ref{eq:halo-model}), one has to model these quantities first. This is done in the remainder of this section. We follow the notation of Ref.~\cite{Barreira:2013xea}, to which we refer the reader for further details of the derivation.

\subsection{Halo mass function}

The halo mass function can be expressed as

\bq\label{eq:mass-function}
&&\frac{{\rm d}n(M)}{{\rm d}{\rm ln}M}{\rm d}{\rm ln}M = \frac{\bar{\rho}_{m0}}{M} f(S){\rm d}S,
\eq
where $S$ denotes the variance of the linear density field filtered on some comoving distance scale $R$,

\bq\label{eq:variance}
S(R) \equiv \sigma^2(R) = 4\pi\int k^2 P_{k, {\rm lin}}\tilde{W}^2\left(k, R\right){\rm d}k.
\eq
Here, $\tilde{W}\left(k, R\right) = 3\left({\rm sin}(kR) - kR{\rm cos}(kR)\right)/\left(kR\right)^3$ is the Fourier transform of the filter function, which we take as a top-hat in real space. The mass enclosed by the filter is given by

\bq\label{eq:mass-overdensity}
M = 4\pi \bar{\rho}_{m0}R^3/3.
\eq
In Eq.~(\ref{eq:mass-function}), $f(S){\rm d}S$ is associated with the fraction of the total mass that resides in haloes whose variances fall within $\left[S, S + {\rm d}S\right]$ (or equivalently, whose masses fall within $\left[M-{\rm d}M, M \right]$) \footnote{Note that for fixed cosmological parameters, the quantities $S$, $R$ and $M$ can be related to one another via Eqs.~(\ref{eq:variance}) and (\ref{eq:mass-overdensity}). In the remainder of the paper, we use these quantities interchangeably when referring to the scale of the haloes.}. 

Motivated by the ellipsoidal collapse of overdense regions, Refs.~\cite{Sheth:1999mn, Sheth:1999su, Sheth:2001dp} proposed the following expression for $f(S)$

\bq\label{eq:first-crossing-ST}
f(S) = A \sqrt{\frac{q}{2\pi}}\frac{\delta_c}{S^{3/2}}\left[1 + \left(\frac{q\delta_c^2}{S}\right)^{-p}\right]{\rm exp}\left[-q\frac{\delta_c^2}{2S}\right], \nonumber \\
\eq
where $\delta_c \equiv \delta_c(z)$ is the critical initial overdensity for a spherical top-hat to collapse at redshift $z$, extrapolated to $z = 0$ with the $\Lambda$CDM linear growth factor \footnote{This extrapolation is done purely  to ensure that the resulting values of $\delta_c$ in the Galileon model can be more readily compared to results from other models in the literature.}. The values of $\delta_c(z)$ for the Cubic and Quartic Galileon models are shown in Table \ref{table:dc-values}. Note that for consistency, $P_{k, {\rm lin}}$ in Eq.~(\ref{eq:variance}) is also the initial power spectrum of the specific model (say the Cubic or the Quartic Galileon models), evolved to $z = 0$ with the $\Lambda$CDM linear growth factor. The choice of parameters $(q,p) = (1,0)$ leads to the Press-Schechter mass function \cite{1974ApJ...187..425P}, whose shape and amplitude are motivated by the spherical (rather than ellipsoidal) collapse of matter overdensities. However, Refs.~\cite{Sheth:1999mn, Sheth:1999su, Sheth:2001dp} found that the choice of parameters $(q,p) = (0.75, 0.30)$ provides a much more accurate description of the mass function measured from N-body simulations of $\Lambda$CDM models. For Galileon gravity models, it is not necessarily true that this choice of $(q,p)$ parameters also results in a good fit to N-body results. In the next section, we recalibrate these two parameters to our simulations of the Cubic and Quartic Galileon models. The normalization constant $A$ is fixed by requiring that $\int f(S){\rm d}S = 1$. The mass function computed using Eqs.~(\ref{eq:mass-function}) and (\ref{eq:first-crossing-ST}) is known as the Sheth-Tormen mass function. 

\subsection{Linear halo bias}

The linear halo bias parameter $b(M)$ \cite{Mo:1995cs} quantifies the difference between the clustering amplitude of haloes of mass $M$ and that of the underlying total dark matter field on large scales ($k \ll 1 h/{\rm Mpc}$),

\bq\label{eq:bias-def}
\delta_{\rm halo}(M) = b(M)\delta_{\rm matter},
\eq
where $\delta_{\rm halo}$ and $\delta_{\rm matter}$ represent, respectively, the density contrast of the distribution of haloes of mass $M$ and of the total dark matter field. Equation (\ref{eq:bias-def}) is only valid when $|\delta_{\rm matter}| \ll 1$, i.e., on large cosmological scales. On smaller scales, where the matter overdensity is larger, higher order terms are needed \cite{Fry:1992vr}.

By following the same steps as in Ref.~\cite{Barreira:2013xea}, one can straightforwardly show that the Sheth-Tormen halo bias is given by

\bq\label{eq:st-halo-bias}
b(M) = 1 + {g(z)}\left(\frac{q{\delta_c^2}/{S} - 1}{\delta_c} + \frac{2p/\delta_c}{1 + \left(q\delta_c^2/S\right)^p}\right),
\eq
with $g(z) = D^{\Lambda \rm CDM}(z = 0) / D^{\rm Model}(z)$, where $D(z)$ is the linear growth factor of a specific model. The latter is defined as $\delta_{\rm matter}(z) = D(z)\delta_{\rm matter}(z_i)/D(z_i)$. Provided the $(q,p)$ parameters are calibrated to fit the mass function, the linear halo bias $b(M)$ should, according to the excursion set theory logic, give automatically a reasonably good fit to the simulation results. This is one of the well-known lessons of Refs.~\cite{Sheth:1999mn, Sheth:1999su, Sheth:2001dp} for the CDM family of models. The results in the next section show that this remains true for the Cubic and Quartic Galileon models.

\subsection{Halo density profiles}

We assume that the radial profile of the dark matter haloes is of the NFW type \cite{Navarro:1996gj} \footnote{Not to be confused with the top-hat profile assumption used in the spherical collapse to obtain the values of the critical density $\delta_c$.}

\bq\label{eq:nfw}
\rho_{\rm NFW}(r) = \frac{\rho_s}{{r}/{r_s}\left[1 + {r}/{r_s}\right]^2},
\eq
where $\rho_s$ and $r_s$ are often called the {\it characteristic density} and the {\it scale radius} of the halo.

The mass of the NFW density profile, $M_{\Delta}$, can be obtained by integrating Eq.~(\ref{eq:nfw}) up to some radius $R_{\Delta}$ (the meaning of the subscript $\Delta$ will become clear later)

\bq\label{eq:mass-nfw}
M_{\Delta} &=& \int_0^{R_{\Delta}}{\rm d}r4\pi r^2 \rho_{\rm NFW}(r) \\ \nonumber
&=& 4\pi \rho_s\frac{R_{\Delta}^3}{c_{\Delta}^3}\left[{\rm ln}\left(1+c_{\Delta} \right) - \frac{c_{\Delta}}{1+c_{\Delta}}\right],
\eq
where we have used the concentration parameter 

\bq\label{eq:c-def}
c_{\Delta} = \frac{R_{\Delta}}{r_s}
\eq
(not to be confused with the $c_i$ parameters in the action of the Galileon model Eq.~(\ref{eq:action})).

In our simulations, the halo mass is defined as

\bq\label{eq:mass-sim}
M_{\Delta} = \frac{4\pi}{3}\Delta\bar{\rho}_{c0}R_{\Delta}^3,
\eq
i.e., $M_{\Delta}$ is the mass enclosed by the comoving radius $R_{\Delta}$, within which the mean density is $\Delta$ times the critical density of the Universe today, $\bar{\rho}_{c0}$. In this paper we consider $\Delta= 200$, but for now let us keep the discussion as general as possible. By combining Eqs.~(\ref{eq:mass-nfw}) and (\ref{eq:mass-sim}), one finds $\rho_s$ as a function of $c_{\Delta}$:

\bq\label{eq:rhos-nfw}
\rho_s = \frac{1}{3}\Delta\bar{\rho}_{c0} c_{\Delta}^3\left[{\rm ln}\left(1+c_{\Delta} \right) - \frac{c_{\Delta}}{1+c_{\Delta}}\right]^{-1}.
\eq
All that is needed to fully specify the NFW profile is to determine the value of $r_s$, which is done by direct fitting to the halo density profiles measured from the simulations. In the literature, however,  it has become more common to specify the concentration-mass relation $c_{\Delta}(M_{\Delta})$, instead of the equivalent values of $r_s$. Previous studies \cite{Bullock:1999he, Neto:2007vq, Maccio':2008xb, Prada:2011jf} have found that the concentration-mass relation is well described by a power law function. The parameters of the power law, however, seem to have a sizeable cosmology dependence, even for different choices of cosmological parameters in $\Lambda$CDM models (see e.g.~Ref.~\cite{Maccio':2008xb}). In the next section, we will see that the $c_{\Delta}(M_{\Delta})$ relation in Galileon models can also be well fitted by a power law, but with fitting parameters that differ considerably from those obtained for $\Lambda$CDM. Having found the $c_{\Delta}(M_{\Delta})$ relation from the simulations, then the NFW density profile becomes completely specified by the mass $M_{\Delta}$ of the halo.

Finally, because what enters Eqs.~(\ref{eq:halo-model-terms}) and (\ref{eq:I-function}) is the Fourier transform of the profiles, $u(k, M)$, and not the profiles themselves, we simply mention that it is possible to show that

\bq\label{eq:nfw-fourier}
u_{\rm NFW}(k, M) &=& \int_0^{R_\Delta} {\rm d}r4\pi r^2\frac{{\rm sin}kr}{kr}\frac{\rho_{\rm NFW}(r)}{M_{\Delta}} \nonumber \\
&=& {4\pi \rho_s r_s^3}\left\{\frac{{\rm sin}\left(kr_s\right)}{M}\left[{\rm Si}\left(\left[1+c_{\Delta}\right]kr_s\right) - {\rm Si}\left( kr_s\right)\right]\right. \nonumber \\
&&\ \ \ \ \ \ \ \ \ \ \ + \left.\frac{{\rm cos}\left(kr_s\right)}{M}\left[{\rm Ci}\left(\left[1+c_{\Delta}\right]kr_s\right) - {\rm Ci}\left( kr_s\right)\right]\right. \nonumber \\
&&\ \ \ \ \ \ \ \ \ \ \  - \left.  \frac{{\rm sin}\left(c_\Delta kr_s\right)}{M\left(1+c_\Delta\right)kr_s}\right\},
\eq
where ${\rm Si}(x) = \int_0^x {\rm d}t{\rm sin}(t)/t$ and ${\rm Ci}(x) = -\int_x^\infty {\rm d}t{\rm cos}(t)/t$. Note that, indeed, $u(k \rightarrow 0,M) \rightarrow 1$, as required.

\section{Results}\label{sec:results}

In this section, we test the predictions of the formulae presented in the last section with the results from N-body simulations of the Cubic \cite{Barreira:2013eea} and Quartic \cite{Li:2013tda} Galileon models. 

\subsection{Summary of the simulations}

\begin{table}
\caption{Summary of the three variants of the Cubic and Quartic Galileon models studied in this paper. All variants have the same background expansion history, but differ in the force law.}
\begin{tabular}{@{}lccccccccccc}
\hline\hline
\\
Model  & \ \ Expansion history & Force law&\ \ 
\\
\hline
\\
Full model                    &\ \  Galileon &  \ \ GR + Screened fifth force &\ \ 
\\
Linear model               &\ \ Galileon &  GR + linear fifth force &\ \ 
\\
$\rm{QCDM}$   &\ \ Galileon &  GR&\ \ 
\\
\hline
\hline
\end{tabular}
\label{table-variants}
\end{table}

The simulations presented in this paper were performed with the {\tt ECOSMOG} code \cite{Li:2011vk}, which is a modified version of the {\tt RAMSES} code \cite{Teyssier:2001cp} that includes extra solvers for the scalar degrees of freedom that appear in modified gravity theories. The code solves the equation of motion of the scalar field by performing Gauss-Seidel iterative relaxations on an adaptively refined grid. The grid is refined whenever the number of particles within a grid cell exceeds some user-specified threshold, $N_{\rm th}$. This ensures that high-density regions are sufficiently well resolved, while saving computational resources in regions where the density is lower. For more details about the code implementation, in particular in Galileon cosmologies, we refer the reader to Refs.~\cite{Li:2011vk, Li:2013nua, Barreira:2013eea, Li:2013tda}.

The results that follow correspond to three variants of the Cubic and Quartic Galileon models. We call the {\it linear model} the one in which the screening mechanism is artificially set to zero, by linearization of the Einstein and the Galileon field equations. The {\it full model} is, as the name indicates, the complete model without any modifications. Comparing these two variants allows one to see the effects of the screening mechanism. The other variant is called {\it QCDM}, and corresponds to a model where there is no contribution from the fifth force, but in which the expansion history is the same as in the other two variants. This variant is used as a base model to identify the modifications to structure formation that arise from the modified force law, excluding those that arise from the modified expansion rate compared to $\Lambda$CDM. The properties of these three variants are summarized in Table \ref{table-variants}. 

The simulations were performed in a box of size $L = 200 {\rm Mpc}/h$, with $N_p = 512^3$ dark matter particles and grid refinement criteria $N_{\rm th} = 8$. For each of the model variants, we have simulated five different realizations of the initial density field, by choosing different random seeds. This allows for statistical averaging, which we use to construct errorbars for the simulation results by measuring the variance within the different realizations.  Although in this paper we show results only for one box size, we note that in Refs.~\cite{Barreira:2013eea, Li:2013tda} the same models were simulated using different box sizes and particle numbers, and gave converged results.

\subsection{Mass function}

\begin{figure*}
	\centering
	\includegraphics[scale=0.390]{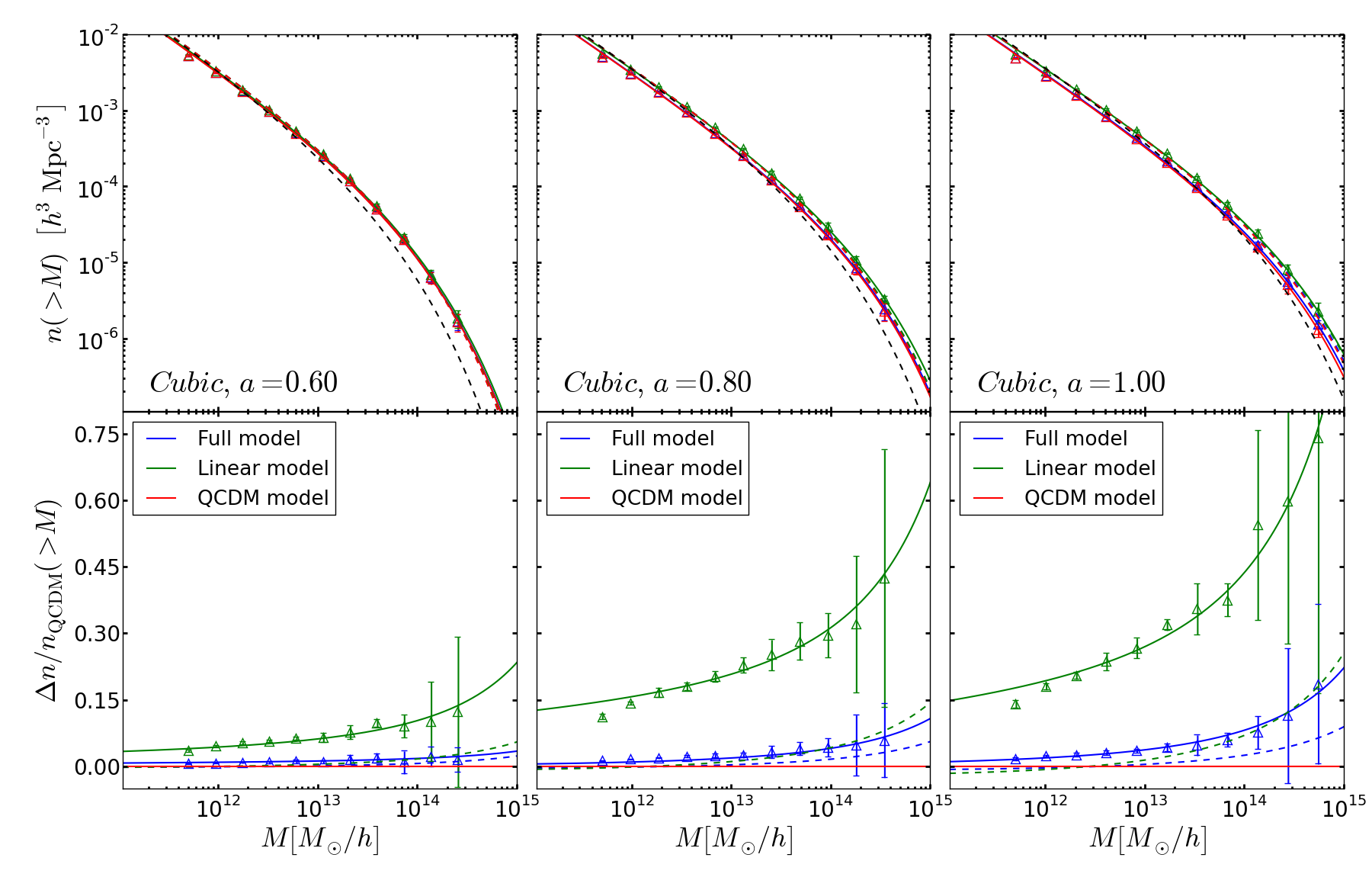}
	\caption{The upper panels show the cumulative mass function of the three variants of the Cubic model at $a = 0.60$, $a = 0.80$ and $a = 1.00$. The triangles with errorbars show the simulation results considering only haloes (and not subhaloes) with mass $M_{200} > 100 M_p$, where $M_p = \Omega_{m0}\bar{\rho}_{c0}L^3/N_p$ is the particle mass. The solid lines correspond to the cumulative mass function predicted using Eqs.~(\ref{eq:mass-function}) and (\ref{eq:first-crossing-ST}), with the best-fitting $(q,p)$ parameters to the simulation results given in Table \ref{table:bf-ap}. The dashed lines are computed in the same way as the solid lines, but with the standard Sheth-Tormen parameter values $(q,p) = (0.75, 0.30)$. For reference, the Sheth-Tormen cumulative mass function for a $\Lambda$CDM model with WMAP9 parameters \cite{Hinshaw:2012fq} is shown by the black dashed curve in the upper panels. The color scheme indicated in the figure applies to the lines and symbols. {In the lower panels, the relative difference of the simulation results w.r.t. the QCDM simulations results is shown, and the relative difference of the analytical predictions is plotted w.r.t. to the analytical predictions of the QCDM model. Also in the lower panels, the solid red and dashed red lines are both zero, by definition.}} 
\label{fig:cmf-cubic}\end{figure*}

\begin{figure*}
	\centering
	\includegraphics[scale=0.390]{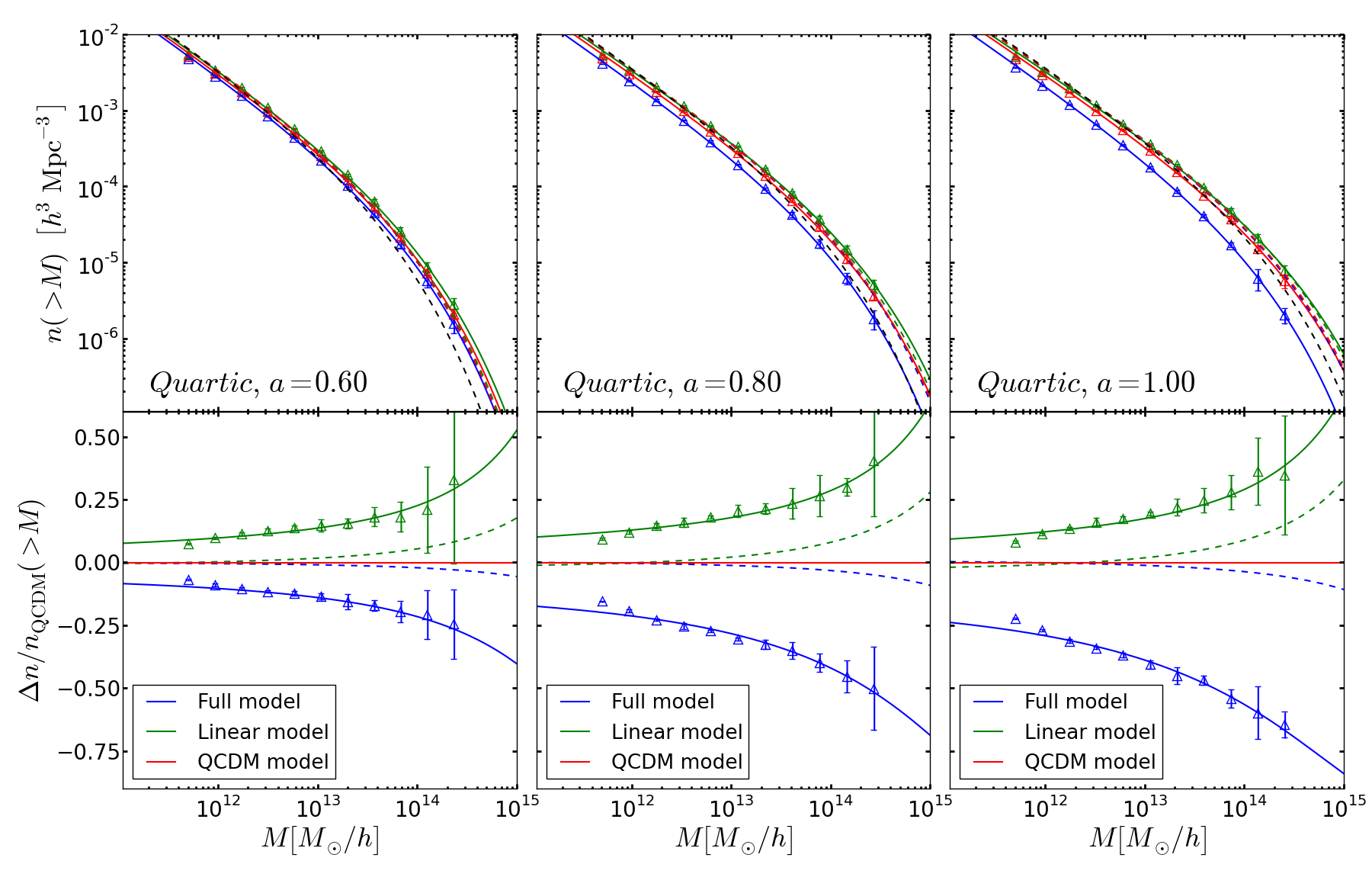}
	\caption{Same as Fig.~\ref{fig:cmf-cubic}, but for the Quartic Galileon model.}
\label{fig:cmf-quartic}\end{figure*}

\begin{table*}
\caption{Best-fitting Sheth-Tormen $(q,p)$ parameters to the simulation results for the variants of the Cubic and Quartic Galileon models at $a = 0.60$, $a = 0.80$ and $a = 1.00$. The uncertainty in the values of $q$ and $p$ is $\Delta_q = 3.5\times10^{-3}$ and $\Delta_p = 1.5\times 10^{-3}$, respectively. These parameters were determined by minimizing the quantity $\sum_i |n^{\rm sims}(>M_i)/n^{\rm ST}(>M_i, q, p) - 1|$, where $n^{\rm sims}$ is the cumulative mass function measured in the simulations, $n^{\rm ST}$ is the analytical result given by the Sheth-Tormen mass function and the index '$i$' runs over the number of bins in the cumulative mass function.}
\begin{tabular}{@{}lccccccccccc}
\hline\hline
\\
Model  & \ \ $a = 0.60$ & $a = 0.80$ & $a = 1.00$&\ \ 
\\
              & \ \ $(q,p)$ & $(q,p)$ & $(q,p)$&\ \ 
\\
\hline
\\
${\rm QCDM}_{\rm Cubic}$                    &\ \  $(0.699, 0.336)$ &  \ \ $(0.727, 0.349)$ & $(0.791, 0.354)$&\ \ 
\\
Cubic Galileon                                             &\ \  $(0.699, 0.334)$ &  \ \ $(0.720, 0.346)$ & $(0.770, 0.349)$&\ \ 
\\
Linearized Cubic Galileon                        &\ \  $(0.685, 0.326)$ &  \ \ $(0.692, 0.308)$ & $(0.734, 0.301)$&\ \ 
\\
\\
${\rm QCDM}_{\rm Quartic}$                    &\ \  $(0.671, 0.339)$ &  \ \ $(0.692, 0.349)$ & $(0.713, 0.354)$&\ \ 
\\
Quartic Galileon                                             &\ \  $(0.713, 0.359)$ &  \ \ $(0.840, 0.389)$ & $(1.024, 0.407)$&\ \ 
\\
Linearized Quartic Galileon                        &\ \  $(0.649, 0.316)$ &  \ \ $(0.671, 0.316)$ & $(0.692, 0.321)$&\ \ 
\\
\hline
\hline
\end{tabular}
\label{table:bf-ap}
\end{table*}

In Figs.~\ref{fig:cmf-cubic} and \ref{fig:cmf-quartic}, we show our results for the cumulative mass function of the Cubic and Quartic Galileon models, respectively. These were obtained with the phase-space friends-of-friends halo finder code {\tt Rockstar} \cite{Behroozi:2011ju}. Throughout, we use $M$ and $M_{200}$ interchangeably to denote halo mass. The symbols with errorbars indicate the simulation results and the dashed lines show the mass function predicted by Eqs.~(\ref{eq:mass-function}) and (\ref{eq:first-crossing-ST}) using the values of $\delta_c$ from Table \ref{table:dc-values} and the standard Sheth-Tormen parameters $(q,p) = (0.75, 0.30)$. One can see that the mass function computed in this way fails to provide a reasonable description of the simulation results. In terms of the relative difference w.r.t.~QCDM, the standard Sheth-Tormen prediction gets the correct qualitative trend, but significantly underestimates the effects of the modifications to gravity seen in the simulations.

It is not completely surprising that the use of the standard Sheth-Tormen parameters $(q,p) = (0.75, 0.30)$ fails in the Galileon model, since these were chosen to fit $\Lambda$CDM simulations \cite{Sheth:1999mn, Sheth:1999su, Sheth:2001dp}. The ellipsoidal collapse motivates a departure from $(q,p) = (1, 0)$ (which corresponds to the spherical collapse case), but the magnitude of this departure is determined by fitting to numerical results. Very crudely, one can say that the fitted $(q,p)$ parameters absorb some of the uncertain details of the nonlinear structure formation, which cannot be accurately described by the ellipsoidal collapse. In models that differ significantly from $\Lambda$CDM, like the Cubic or Quartic Galileon models, it is to be expected that the specifics of the ellipsoidal collapse should also be different. In practice, this translates into different values for the $(q,p)$ parameters. The solid lines in Figs.~\ref{fig:cmf-cubic} and \ref{fig:cmf-quartic} show the mass function predicted using the values of $\delta_c$ from Table \ref{table:dc-values} and the best-fitting $(q,p)$ parameters to the simulation results. The latter were determined for each variant of the Cubic and Quartic Galileon models at $a = 0.6$, $a = 0.8$ and $a = 1$. Their values are shown in Table \ref{table:bf-ap}. By allowing $(q,p)$ to differ from the standard values, one sees that the analytical predictions of Eqs.~(\ref{eq:mass-function}) and (\ref{eq:first-crossing-ST}) can actually provide an extremely good fit to the simulation results in the entire mass range probed. Similarly, Ref.~\cite{Wyman:2013jaa} has found that, by appropriately adjusting the values of parameters in the Tinker mass function \cite{Tinker:2008ff}, then the latter can match the simulation results of a model that also employs the Vainshtein effect but with a $\Lambda$CDM expansion history.

In the case of the Cubic model, one sees that the screening mechanism works well at suppressing the enhancement in the number density of haloes. For instance, the linear variant predicts an enhancement in the number density of haloes with $M \sim 10^{14} M_{\odot}/h$ at $a = 1$ of about $45\%$, whereas in the case of the full model, in which the screening is at play, the enhancement is smaller than $10\%$. On the other hand, in the case of the Quartic model, the overall weakening of gravity in the full variant (cf.~Fig.~(\ref{fig:Geffmap})) leads to a significant suppresion in the number density of collapsed objects. In particular, haloes with $M \sim 10^{14} M_{\odot}/h$ are $\sim 50\%$ less abundant compared to QCDM. In the case of the linearized variant, the same massive haloes are $\sim 30\%$ more abundant w.r.t. QCDM \footnote{See also Refs.~\cite{Hellwing2009, Hellwing2010} for N-body simulations of a phenomenological model with a Yukawa-like fifth force, for which similar results are found.}.

Before proceeding, a comment should be made about the definition of halo mass in the simulations and in the analytical formulae presented in Sec.~\ref{sec:halo-model-formulae}. Assuming mass conservation, Eq.~(\ref{eq:mass-overdensity}) can be associated with the virial mass of the halo, whose definition differs in different models. In this paper, we are comparing the mass $M$ of Eq.~(\ref{eq:mass-overdensity}) with the values of $M_{200}$ measured from the simulations. One does not expect these two mass definitions to be exactly the same, but nor would one expect them to differ significantly. These ambiguities in the mass definition can, anyway, be absorbed in the fitted values of the Sheth-Tormen $(q,p)$ parameters. We expect these fitting parameters to slightly change with different mass definitions. However, note that this is also the case for the $\Lambda$CDM model, and is not peculiar to Galileon gravity.

\subsection{Linear halo bias}

\begin{figure*}
	\centering
	\includegraphics[scale=0.390]{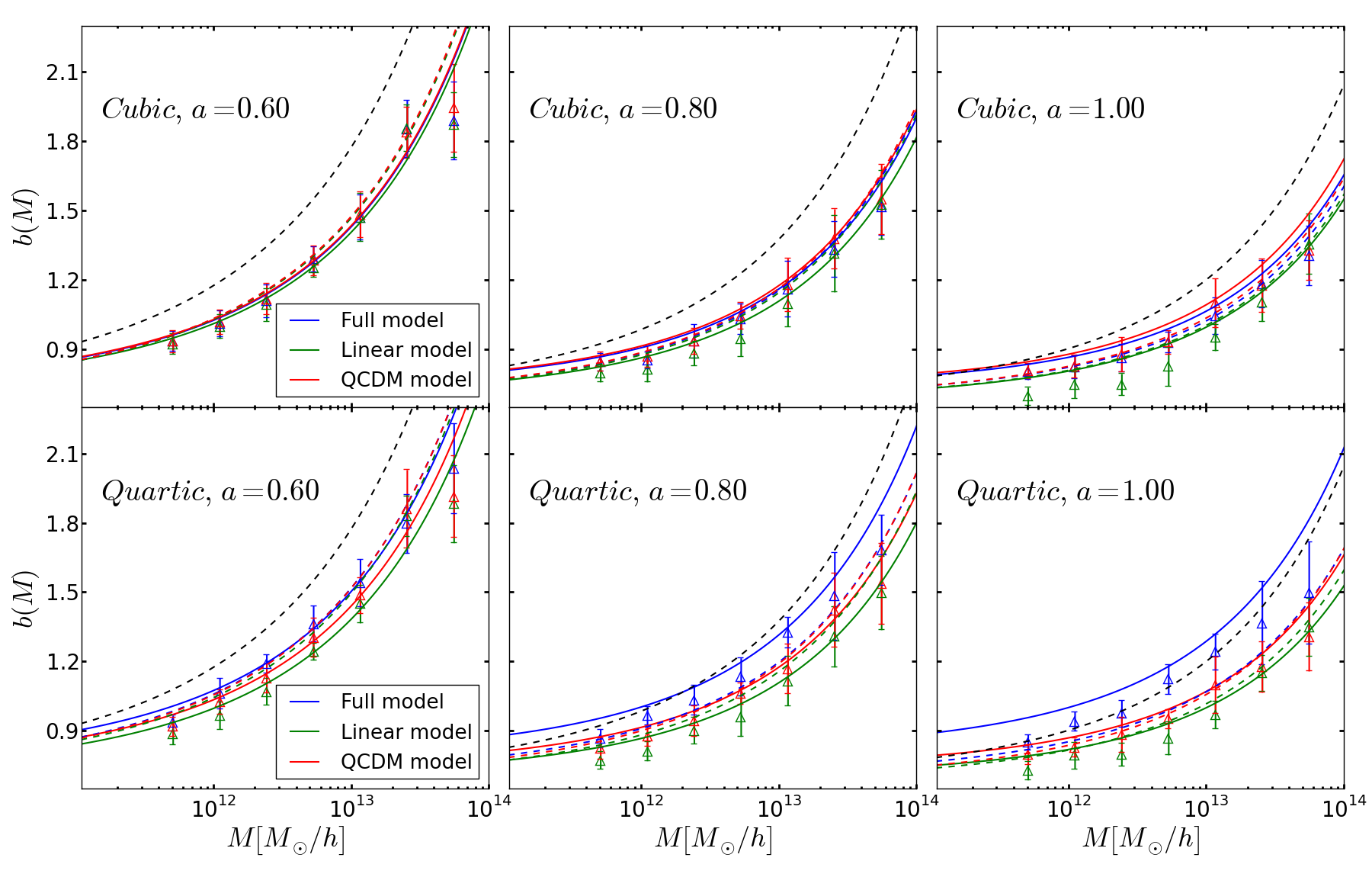}
	\caption{Linear halo bias of the three variants of the Cubic (upper panels) and Quartic Galileon (lower panels) models at $a = 0.60$, $a = 0.80$ and $a = 1.00$. The triangles with errorbars show the simulation results considering only haloes (and not subhaloes) with mass $M_{200} > 100 M_p$, where $M_p = \Omega_{m0}\bar{\rho}_{c0}L^3/N_p$ is the particle mass. The solid and dashed lines correspond to the linear halo bias parameter of Eq.~(\ref{eq:st-halo-bias}) computed with the $(q,p)$ parameters from Table \ref{table:bf-ap} and $(q,p) = (0.75, 0.30)$, respectively. The linear halo bias for a $\Lambda$CDM model with WMAP9 parameters \cite{Hinshaw:2012fq} is shown by the black dashed lines. The color scheme indicated in the figure applies to the lines and symbols.}
\label{fig:hb}\end{figure*}

In our simulations, we measure the halo bias by evaluating the ratio

\bq\label{eq:bias-def}
b(k,M) = \frac{P_{\rm hm}(k,M)}{P(k)},
\eq
where $P(k)$ is the total matter power spectrum and  $P_{\rm hm}(k,M)$ is the halo-matter cross spectrum for haloes of mass $M$. We used a {\it Delaunay Tessellation} field estimator code \cite{cv2011, sv2000} to obtain the halo and matter density fields from which we computed these power spectra. In the numerator of Eq.~(\ref{eq:bias-def}), we consider the cross power spectrum, rather than the halo-halo counts power spectrum, to reduce the impact of shot noise on our measurements. Our estimate for the linear halo bias is given by the asymptotic value of $b(k, M)$ on large scales (small $k$). The result is shown in Fig.~\ref{fig:hb} for the Cubic (upper panels) and Quartic (lower panels) Galileon models.  {In Fig.~\ref{fig:hb}, one sees that Eq.~(\ref{eq:st-halo-bias}) provides a good description of the linear halo bias seen in the simulations if one uses the $(q,p)$ parameters that best-fit the mass function of the simulations (solid lines). This shows that the excursion set theory approach and the steps involved in the derivation of Eq.~(\ref{eq:st-halo-bias}) are still valid in the Cubic and Quartic Galileon models. However, the use of the best-fitting $(q,p)$ parameters does not lead to a significant improvement over the use of the standard Sheth-Tormen values, $(q,p) = (0.75, 0.30)$ in matching the simulation results. The linear halo bias seems to be less sensitive than the halo mass function to the exact choice of $(q,p)$. This can be understood as the linear halo bias is computed as the ratio of two mass functions \cite{Mo:1995cs, Barreira:2013xea}, and consequently, some of the dependence on the values of $(q,p)$ cancels to some extent. Note that despite the weaker sensitivity to the exact choice of the Sheth-Tormen parameters, these must still differ from the Press-Schechter limit $(q,p) = (1,0)$, which is known to fail to reproduce the results from N-body simulations \cite{Sheth:1999mn, Sheth:1999su, Sheth:2001dp}.}

\subsection{Halo occupation distribution analysis}

\begin{figure*}
	\centering
	\includegraphics[scale=0.390]{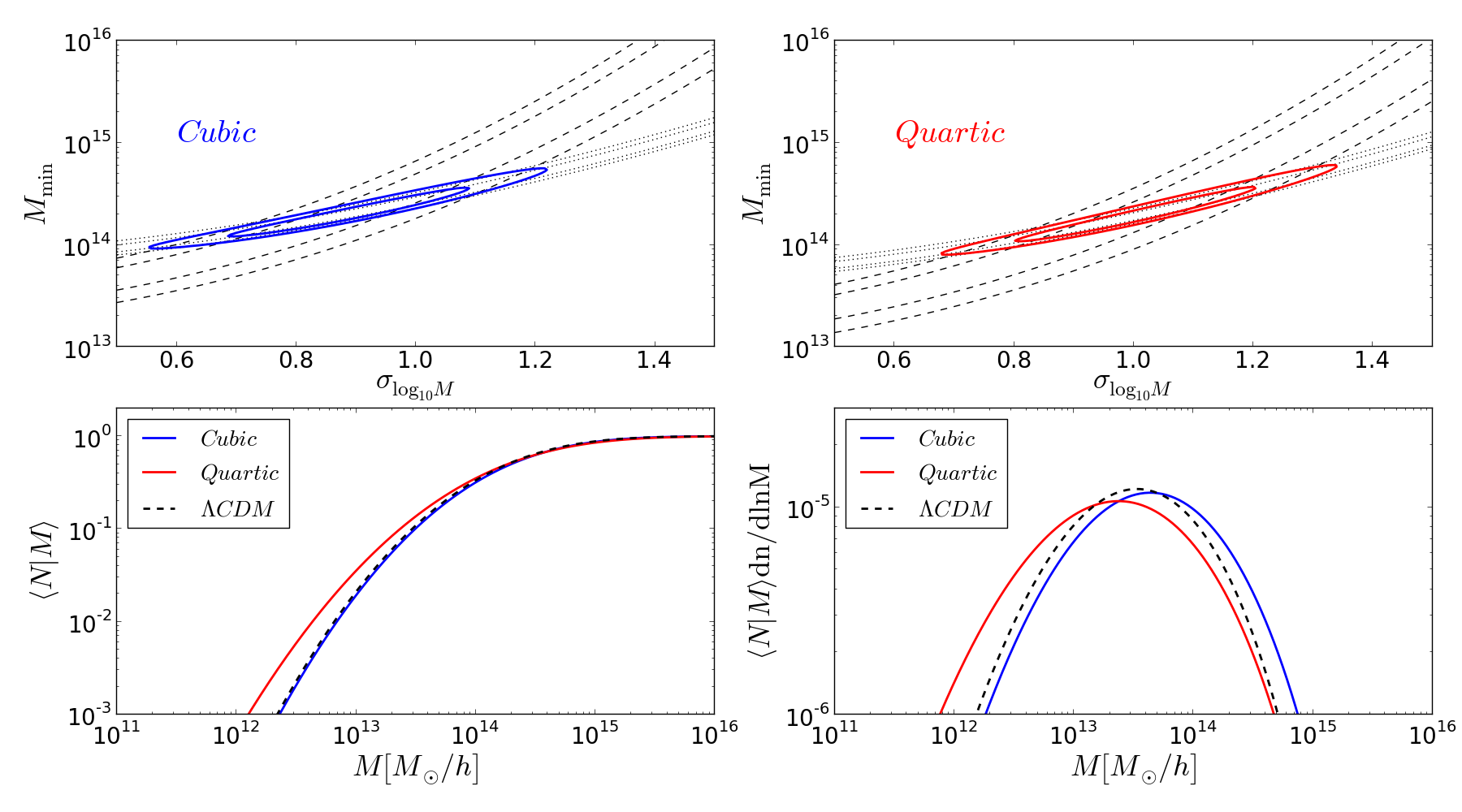}
	\caption{{The upper panels show the 68\% and 95\% confidence contours on the $M_{\rm min}$ and $\sigma_{{\rm log}_{10}M}$ parameters of Eq.~(\ref{eq:hod-model}) obtained using Eq.~(\ref{eq:hod-chi2}) for the full variants of the Cubic (left) and Quartic (right) Galileon models. The black dashed and dotted contours indicate the constraints derived by using only $\chi^2_{b_g}$ and only $\chi^2_{n_g}$ in Eq.~(\ref{eq:hod-chi2}), respectively. The solid contours show the combined constraints. The lower left and lower right panels show, respectively, the best-fitting $\left<N|M\right>$ and $\left<N|M\right> {\rm d}n/{\rm dln}M$ for the full Cubic (blue) and Quartic (red) Galileon models, and $\Lambda$CDM with WMAP9 parameters \cite{Hinshaw:2012fq} (black dashed). The quantity plotted in the lower right panel shows the contribution from haloes of different mass to the galaxy number density.}}
\label{fig:hod}\end{figure*}

As indicated by the values of $\sigma_8 \sim 1$ in Table \ref{table:table-max}, the amplitude of the linear matter power spectrum in the Cubic and Quartic Galileon models is higher than in standard $\Lambda$CDM models, for which $\sigma_8 \sim 0.82$ \cite{Hinshaw:2012fq}. Consequently, it is interesting to investigate if the enhanced clustering power in the Galileon models is still consistent with the observed large scale clustering of the host haloes of Luminous Red Galaxies (LRGs) of the SDSS DR7 \cite{Reid:2009xm}. The screening mechanism could potentially suppress part of the enhancement, but Refs.~\cite{Barreira:2013eea, Li:2013tda} have shown that the impact of the Vainshtein effect is negligible on sufficiently large scales ($k \lesssim 0.1 h/{\rm Mpc}$). On the other hand, the result of Fig.~\ref{fig:hb} shows that massive haloes in Galileon cosmologies can be less biased than in $\Lambda$CDM, which effectively suppresses the halo power spectrum. As a result, a robust comparisson between theory and observations requires an exploration of this degeneracy between the enhanced linear growth of structure and the lower halo bias parameter. We carry on such an exploration by performing a halo occupation distribution (HOD) analysis of LRG clustering.

In the HOD formalism, one asks what is the probability distribution $P(N,M)$ that a dark matter halo of mass $M$ contains $N$ galaxies. The HOD models are typically parametrized by the mean of their distribution, $\left<N|M\right>$, which can be separated into the mean number of {\it central} and {\it satellite} galaxies that reside in haloes of mass $M$ \cite{Kravtsov:2003sg}. For simplicity, and since independent HOD studies have suggested that the satellite fraction is small for LRGs \cite{Wake:2008mf, Zheng:2008np, Sawangwit:2009bg}, we neglect the contribution from satellite galaxies and assume that the haloes can either host one LRG (the central) or none at all. Our aim is to determine if it is possible, in the Cubic and Quartic Galileon cosmologies, to realistically populate the dark matter haloes with LRGs in order to reproduce the observed clustering amplitude and galaxy number density. We parametrize the HOD as

\bq\label{eq:hod-model}
\left<N|M\right> = \frac{1}{2}\left[1 + {\rm erf}\left(\frac{{\rm log}_{10}\left(M/M_{\rm min}\right)}{\sigma_{{\rm log}_{10}M}}\right)\right],
\eq
where $M_{\rm min}$ and $\sigma_{{\rm log}_{10}M}$ are the HOD parameters. These can be constrained by constructing the following $\chi^2$ quantity

\bq\label{eq:hod-chi2}
\chi^2 = \chi^2_{b_g} + \chi^2_{n_g} = \frac{\left(b_g -\bar{b}_g\right)^2}{\Delta b_g} + \frac{\left(n_g - \bar{n}_g\right)^2}{\Delta n_g},
\eq
where

\bq
\label{eq:ng} n_g &=& \int{\rm d}M \frac{\rm{d}n}{\rm{d}M}\left<N|M\right>, \\
\label{eq:bg} b_g &=& \int{\rm d}M \frac{\rm{d}n}{\rm{d}M}\left<N|M\right>b(M),
\eq
are the number density and effective linear bias parameter of the galaxies, respectively. The likelihood of $M_{\rm min}$ and $\sigma_{{\rm log}_{10}M}$ is then $\mathcal{P} \propto {\rm exp}\left[-\chi^2/2\right]$.

In Eq.~(\ref{eq:hod-chi2}),  $\bar{n}_g$ and $\bar{b}_g$ are, respectively, the number density and galaxy bias of the LRG sample presented in Ref.~\cite{Reid:2009xm}, which the HOD model should reproduce. We take $\bar{n}_g = 4 \times 10^{-5} h^3/{\rm Mpc}^3$, which corresponds roughly to $\int n(z) {\rm d}z$, where $n(z)$ is the redshift dependence of the observed galaxy number density (see Fig.~1 of Ref.~\cite{Reid:2009xm}). The value of $\bar{b}_g$ can be inferred from the ratio $R$ of the amplitudes of the observed LRG host halo power spectrum (in redshift space) and the theoretical linear prediction for each model (in real space). To first approximation we can write:

\bq\label{eq:relamp}
R \equiv \frac{P_{k,\rm{LRG}}^{s}(z_{\rm eff})}{P^{r}_{k, \rm lin}(z_{\rm eff})} = \left(\frac{\bar{b}_g}{1.85}\right)^2\left[1 + \frac{2f}{3\bar{b}_g} + \frac{f^2}{5\bar{b}_g^2}\right],
\eq
where $f = {\rm dln}D/{\rm dln}a$ is the logarithmic derivative of the linear growth factor at $z_{\rm eff} = 0.313$, which is the effective redshift of the LRG sample. On the RHS of Eq.~(\ref{eq:relamp}), the term within squared brackets approximately describes the boost in the real space power spectrum caused by the peculiar velocities of galaxies on large scales. The  $\bar{b}_g^2$ factor accounts for the shift in the power due to the galaxy bias. Although $\bar{b}_g$ is the bias of the LRGs, the method used in Ref.~\cite{Reid:2009xm} effectively leads to a normalization of the LRG host halo power spectrum with a factor $(1.85)^{-2}$ (see their Erratum \cite{Reid11112011}). For the Cubic and Quartic Galileon models, we have that $f_{\rm{Cubic}} \approx f_{\rm{Quartic}} \approx 0.75$ and $R_{\rm{Cubic}} \approx R_{\rm{Quartic}} \approx 1.10$ (see Fig.~1 of Ref.~\cite{Barreira:2013xea}). Solving Eq.~(\ref{eq:relamp}) yields $\bar{b}_g \approx 1.68$. For reference, in a $\Lambda$CDM model with WMAP9 parameters \cite{Hinshaw:2012fq}, one has $f \approx 0.66$ and $R \approx 1.40$ which leads to $\bar{b}_g \approx 1.96$. In Eqs.~(\ref{eq:ng}) and (\ref{eq:bg}), we use the calibrated Sheth-Tormen formulae for the mass function and linear halo bias at $a = 0.80$, which is sufficiently close to $a_{\rm eff} = 1/(1 + z_{\rm eff}) \approx 0.76$. We assume fractional errors of $10\%$ and $5\%$ on the number density and galaxy bias, respectively, i.e.,  $\Delta n_g = 0.1\bar{n}_g$ and $\Delta b_g = 0.05\bar{b}_g$. We have checked that our results do not depend on these assumptions for the size of the errors.

The constraints on the parameters $M_{\rm min}$ and $\sigma_{{\rm log}_{10}M}$ for the full Cubic and Quartic variants are shown in the upper panels of Fig.~\ref{fig:hod}. The dashed and dotted contours show the confidence regions obtained by considering only $\chi^2_{b_g}$ or $\chi^2_{n_g}$ in Eq.~(\ref{eq:hod-chi2}), respectively. The fact that the two contours overlap means that there are some LRG HODs that can match both the observed large scale clustering amplitude and number density. The best-fitting HOD models are shown in the lower left panel of Fig.~\ref{fig:hod}. We also show the best-fitting HOD for $\Lambda$CDM. It is remarkable that the Cubic and $\Lambda$CDM models predict almost the same HOD. This shows that the boost in the linear matter power spectrum can be compensated by the modifications to the halo abundance and linear bias in the Cubic model to preserve the way the LRGs populate the dark matter haloes. In the case of the Quartic model, the lower amplitude of the halo mass function (c.f.~Fig.~\ref{fig:cmf-quartic}) and the higher linear halo bias (c.f.~Fig.\ref{fig:hb}) make the HOD extend towards slightly lower halo masses. The lower right panel of Fig.~\ref{fig:hod} shows the halo mass function weighted by the best-fitting HOD model, $\left<N|M\right> {\rm d}n/{\rm dln}M$. The latter peaks at $M \sim [2-4] \times 10^{13} M_{\odot}/h$ and predicts a negligible fraction of LRGs residing in haloes with mass $< 10^{12} M_{\odot}/h$, for all the models. Note that otherwise this would represent an observational tension since LRGs have stellar masses that are typically $> 10^{11} M_{\odot}/h$ \cite{Almeida:2007ef, Banerji:2009pv, Baugh:2006pf}, and are not expected to reside in dark matter haloes whose mass is comparable to theirs.

From the above analysis, we therefore conclude that it is unlikely that the Cubic and Quartic Galileon models are in tension with the large-scale galaxy distribution. This conclusion is contrary to the concerns raised in Refs.~\cite{Barreira:2012kk, Barreira:2013jma, Barreira:2013xea}. These arose due to a misunderstanding of the renormalizations made to the amplitude of the LRG power spectrum reported in \cite{Reid:2009xm}, which led to the claim that the LRG host haloes in Galileon cosmologies would have to be essentially unbiased for the models to match the observations. This would lead to unrealistic HOD models. However, in the Erratum \cite{Reid11112011} of Ref.~\cite{Reid:2009xm} it is clearly explained that the reported measurements correspond to a power spectrum whose amplitude is actually $1.85^2$ times smaller than the true clustering amplitude of the LRGs. As a result, and as shown in the above HOD analysis, the LRGs in Galileon cosmologies are indeed significanly biased (albeit less than in $\Lambda$CDM), and such effective galaxy bias parameter can be obtained with realistic halo occupation distributions.

\subsection{Concentration-mass relation}

\begin{figure*}
	\centering
	\includegraphics[scale=0.390]{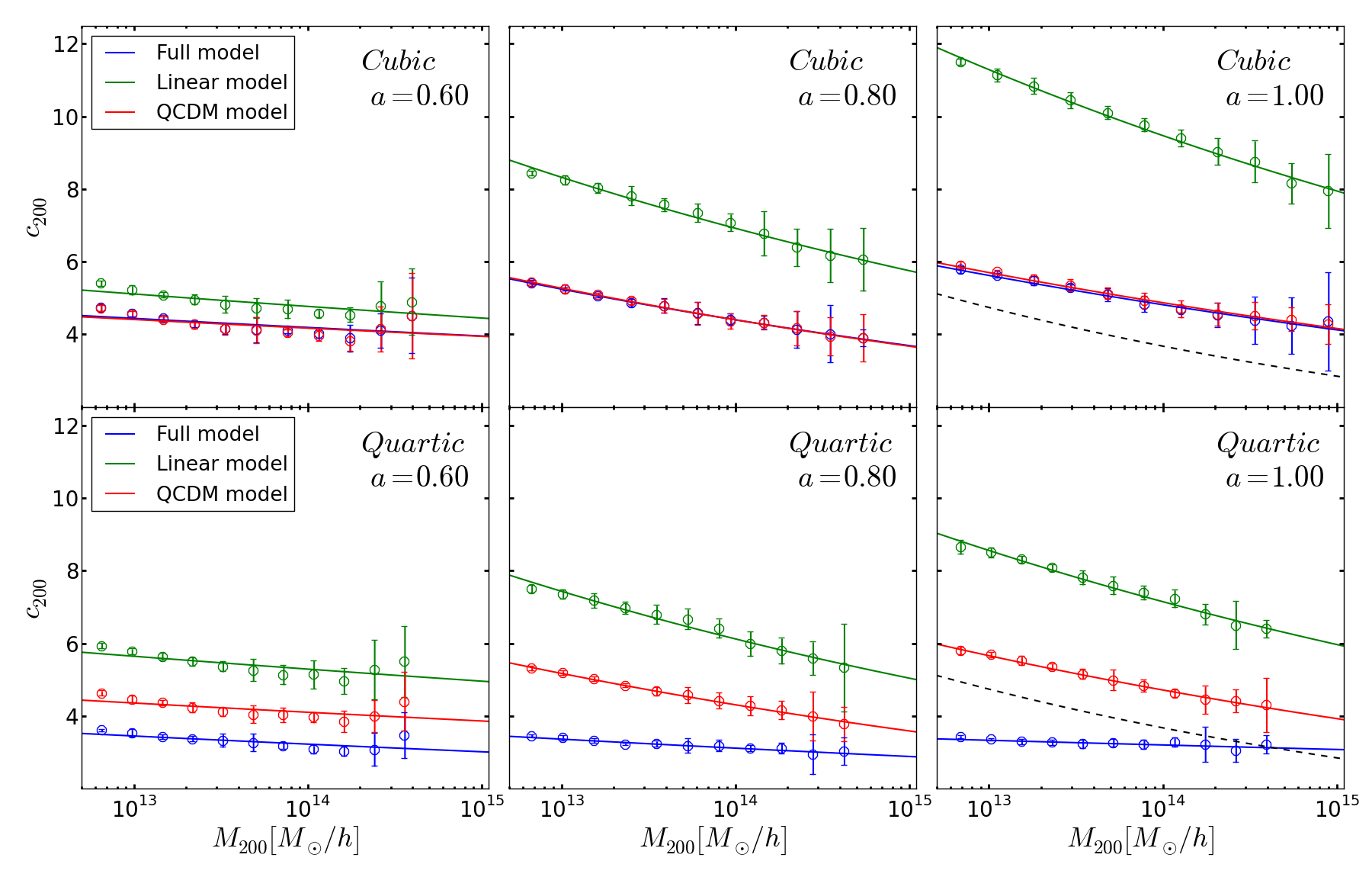}
	\caption{Halo concentration-mass relation, $c_{200}(M_{200})$, of the three variants of the Cubic (upper panels) and Quartic (lower panels) Galileon models, for $a = 0.60$, $a = 0.80$ and $a = 1.00$. The circles with errorbars show the simulation results considering haloes (and not subhaloes) with $M_{200} > 1000 M_p$, where $M_p = \Omega_{m0}\bar{\rho}_{c0}L^3/N_p$ is the particle mass. The solid lines show the best-fitting power laws from Table \ref{table:bf-alphabeta}. For comparison, in the $a = 1.00$ panels, we also show the fit found in Ref.~\cite{Maccio:2008xb} for a $\Lambda$CDM model with the WMAP5 parameters \cite{Komatsu:2008hk}. The color scheme indicated in the figure applies to the lines and symbols.}
\label{fig:c-m}\end{figure*}

\begin{table*}
\caption{Best-fitting $(\alpha, \beta)$ parameters in the parametrization ${\rm log}_{10}(c_{200}) = \alpha + \beta{\rm log}_{10}\left(M_{200} / \left[10^{12} M_{\odot}/h\right]\right)$ to the simulation results for the variants of the Cubic and Quartic Galileon models at $a = 0.60$, $a = 0.80$ and $a = 1.00$. The uncertainty in the values of $\alpha$ and $\beta$ is $\Delta_{\alpha} = \Delta_{\beta} = 0.001$. These parameters were determined by minimizing the quantity $\sum_i |c_{200}^{\rm sims}(M_i)/c_{200}^{\rm param}(M_i, \alpha, \beta) - 1|$, where $c_{200}^{\rm sims}(M_i)$ is the concentration measured in the simulations, $c_{200}^{\rm param}(M_i, \alpha, \beta)$ is the concentration given by the parametrization and the index '$i$' runs over the number of mass bins.}
\begin{tabular}{@{}lccccccccccc}
\hline\hline
\\
Model  & \ \ $a = 0.60$ & $a = 0.80$ & $a = 1.00$&\ \ 
\\
              & \ \ $(\alpha,\beta)$ & $(\alpha,\beta)$ & $(\alpha,\beta)$&\ \ 
\\
\hline
\\
${\rm QCDM}_{\rm Cubic}$                    &\ \  $(0.670, -0.024)$ &  \ \ $(0.801, -0.078)$ & $(0.825, -0.068)$&\ \ 
\\
Cubic Galileon                                             &\ \  $(0.674, -0.025)$ &  \ \ $(0.797, -0.076)$ & $(0.818, -0.067)$&\ \
\\
Linearized Cubic Galileon                        &\ \  $(0.740, -0.030)$ &  \ \ $(1.001, -0.080)$ & $(1.129, -0.076)$&\ \ 
\\
\\
${\rm QCDM}_{\rm Quartic}$                    &\ \  $(0.667, -0.026)$ &  \ \ $(0.794, -0.079)$ & $(0.833, -0.079)$&\ \ 
\\
Quartic Galileon                                             &\ \  $(0.569, -0.029)$ &  \ \ $(0.562, -0.033)$ & $(0.542, -0.017)$&\ \ 
\\
Linearized Quartic Galileon                        &\ \  $(0.781, -0.028)$ &  \ \ $(0.956, -0.084)$ & $(1.011, -0.078)$&\ \ 
\\
\hline
\hline
\end{tabular}
\label{table:bf-alphabeta}
\end{table*}

In Fig.~\ref{fig:c-m}, we show the concentration-mass relation, $c_{200}(M_{200})$, measured in the simulations of the Cubic (upper panels) and Quartic Galileon models (lower panels). {We have checked that the haloes in the simulations are well described by the NFW profile, Eq.~(\ref{eq:nfw}), for all model variants and epochs.} The values of $c_{200}$ were obtained via Eq.~(\ref{eq:c-def}), by using the values of $R_{200}$ and $r_s$ determined by the {\tt Rockstar} code \cite{Behroozi:2011ju}. The simulation results are well fitted by a power law ${\rm log}_{10}(c_{200}) = \alpha + \beta{\rm log}_{10}\left(M_{200} / \left[10^{12} M_{\odot}/h\right]\right)$, with the best-fitting parameters shown in Table \ref{table:bf-alphabeta}. In the Galileon models, one encounters the standard picture that halo concentrations tend to increase with time for fixed mass, and tend to  decrease with halo mass at a given epoch \cite{MunozCuartas:2010ig, Ludlow:2013vxa}. The exact mass and time dependence, however, differs within the variants of the Cubic and Quartic Galileon models. In the $a = 1.00$ panels, we also show a similar power law fitted by Ref.~\cite{Maccio:2008xb} to simulation results of $\Lambda$CDM models with WMAP5 parameters \cite{Komatsu:2008hk}.

In the case of the Cubic Galileon model, one sees that the Vainshtein mechanism is extremely efficient in restraining the modifications to gravity from having an impact on the concentrations of the haloes. The values of $c_{200}$ in the full and in the QCDM variants of the Cubic model are essentially indistinguishable over the mass range probed by the simulations. This is because by the time the modifications to gravity occur, $a \gtrsim 0.6$, the Vainshtein radius of the haloes is larger than the haloes themselves (see e.~g.~Ref.~\cite{Barreira:2013eea}). The Vainshtein radius is the distance from a given matter source below which the fifth force gets suppressed. At $a = 1$, the haloes of all the variants of the Cubic model are more concentrated than in standard $\Lambda$CDM. The reason for this can be traced back to the fact that the haloes in the Cubic models form earlier than in $\Lambda$CDM (cf.~Fig.~\ref{fig:cmf-cubic}). This makes them to be more concentrated since they formed at an epoch when the background matter density was higher. The same reasoning can also be used to explain why the halo concentrations are higher in the linear variant w.r.t.~the QCDM variant. In this case, however, the deepening of the gravitational potential at late times (cf.~Fig.~\ref{fig:Geffmap}) in the linear variant is also expected to play a significant role (see also Refs.~\cite{Hellwing2010b, Hellwing:2008qf}).

The picture in the full Quartic model differs significantly because of the overall weakening of gravity in regions of high density. Following the above reasoning, haloes of a given mass form later in the full Quartic model, which leads to a lower concentration compared to any other variant of the model. The values of $c_{200}$ also barely evolve with time in the full Quartic Galileon model. This can be due to the fact that the gravitational potential inside haloes in the full Quartic model becomes shallower at late times (cf.~Fig.~\ref{fig:Geffmap}).  In fact, the fitting parameter $\alpha$ even decreases with time. Additionally, the mass dependence of the concentration is much shallower than in any other variant, including those of the Cubic Galileon model. Comparing with $\Lambda$CDM at $a = 1$, the full Quartic Galileon model predicts lower halo concentrations, although the difference becomes smaller with increasing halo mass.

Before proceeding, note that since the effects of the fifth force are not felt inside the haloes in the full Cubic Galileon, then our current knowledge about the baryonic processes that are relevant for galaxy formation should prevail \footnote{See, for instance, Ref.~\cite{Sakstein:2013pda} for a study of stellar oscillations in models of modified gravity that employ the chameleon screening mechanism \cite{Khoury:2003rn, Mota:2006fz}.}. As a result, it should be more or less straightforward to implement semi-analytical models of galaxy formation in Cubic Galileon cosmologies. The same, however, does not apply to the Quartic Galileon model.

\subsection{Halo model matter power spectrum}

\begin{figure*}
	\centering
	\includegraphics[scale=0.390]{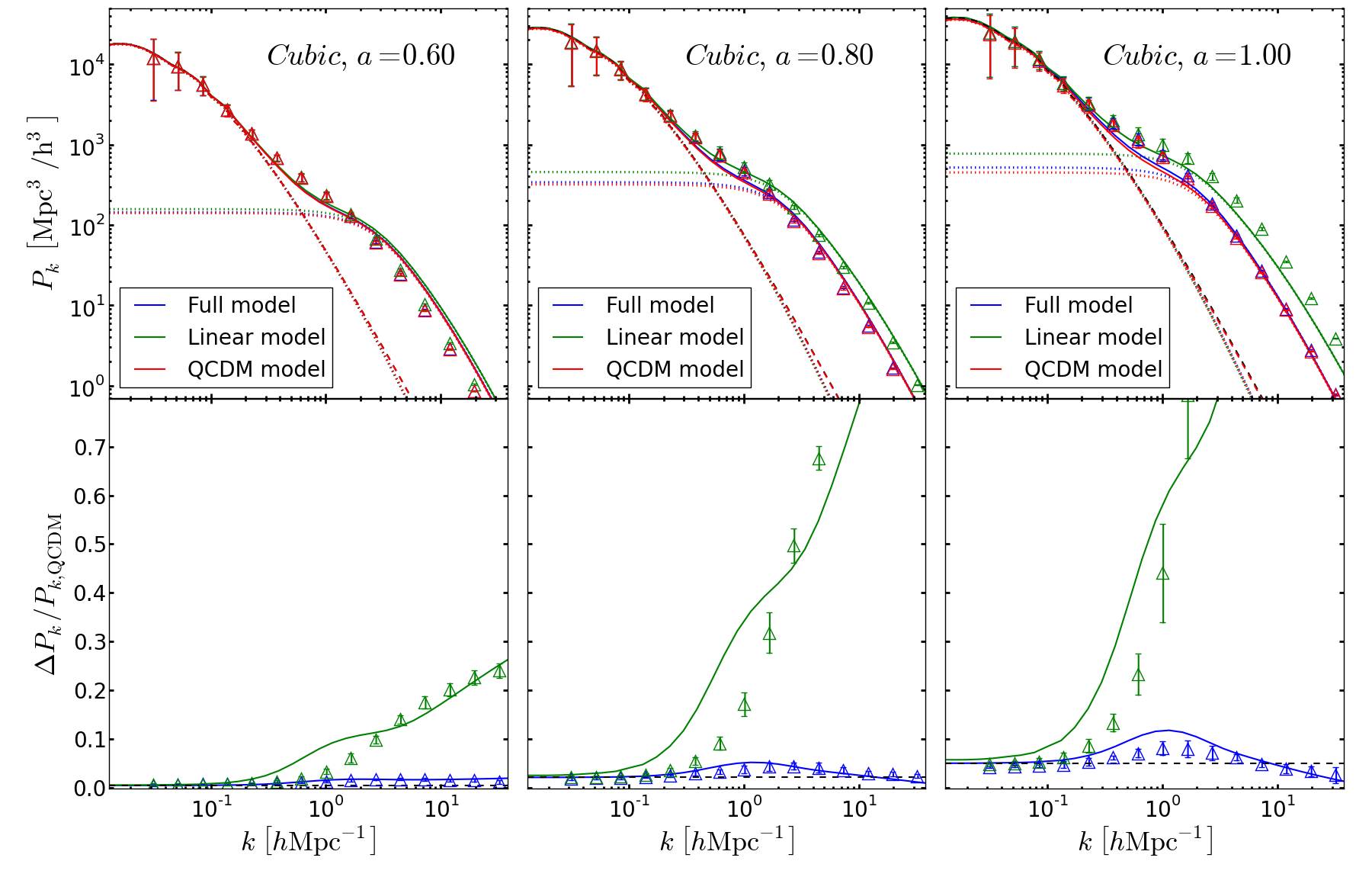}
	\caption{The upper panels show the nonlinear matter power spectrum of the three variants of the Cubic Galileon model, at $a = 0.60$, $a = 0.80$ and $a = 1.00$. The lower panels show the relative difference w.r.t.~QCDM. The triangles with errorbars show the simulation results. The dashed black and dashed red lines show the linear theory prediction for the Cubic Galileon model and its QCDM variant, respectively (these two curves are practically indistinguishable in the upper panels). The solid lines show the nonlinear matter power spectrum in the halo model obtained using Eqs.~(\ref{eq:halo-model}), (\ref{eq:halo-model-terms}) and (\ref{eq:I-function}). The two sets of dotted lines show the contributions from the 1-halo and 2-halo terms. The former is shown by the lines that approach a constant value at small $k$; the latter is shown by the lines that coincide with the linear theory lines at small $k$. The color scheme indicated in the figure applies to the lines and symbols. {In the lower panels, the relative difference of the simulation results w.r.t. the QCDM simulations results is shown, and the relative difference of the analytical predictions is plotted w.r.t. the QCDM analytical predictions.}}
\label{fig:pk-cubic}\end{figure*}

\begin{figure*}
	\centering
	\includegraphics[scale=0.390]{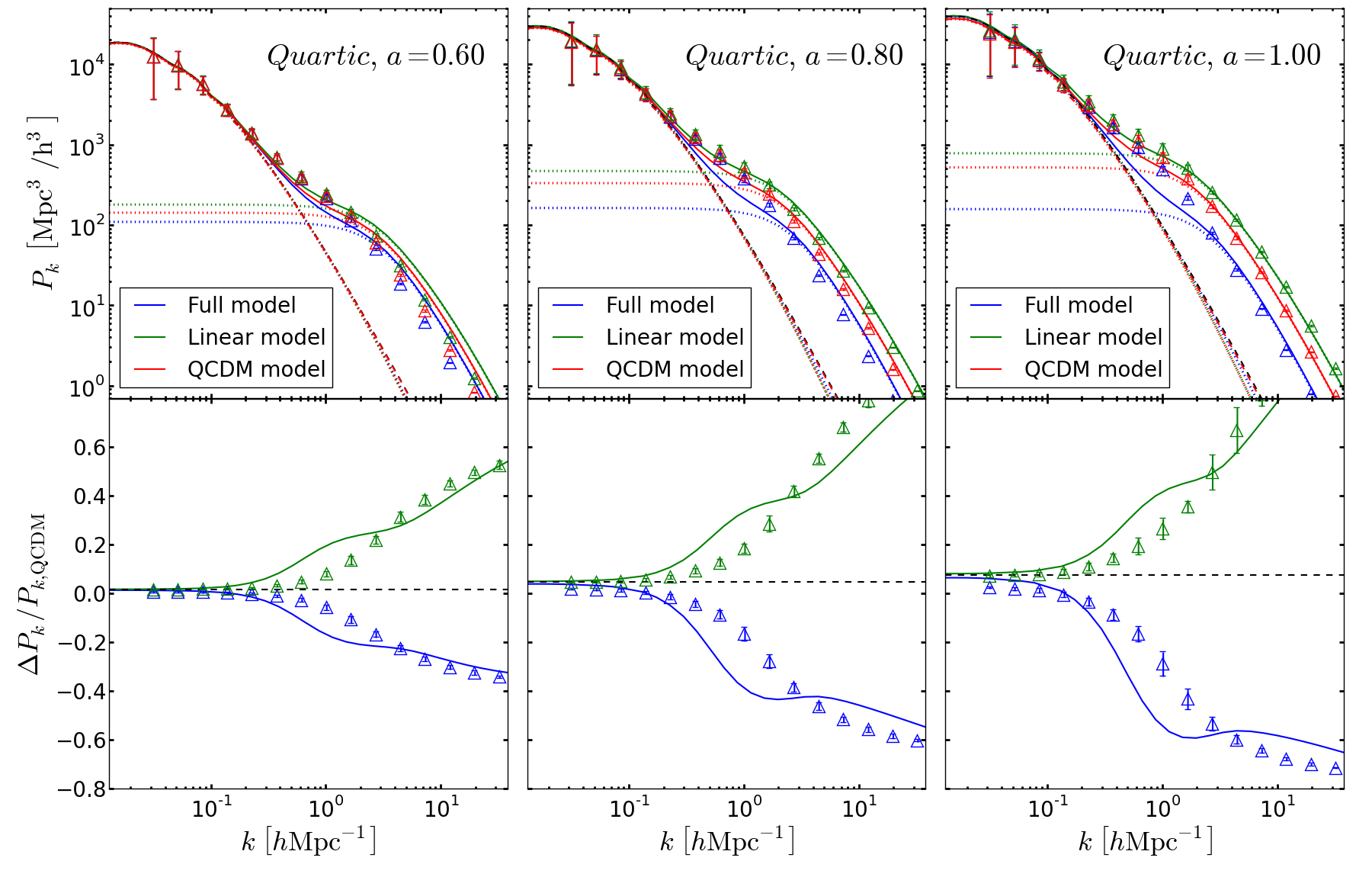}
	\caption{Same as Fig.~\ref{fig:pk-cubic}, but for the Quartic Galileon model.}
\label{fig:pk-quartic}\end{figure*}

Our results for the nonlinear matter power spectrum of the Cubic and Quartic Galileon models are shown in Figs.~\ref{fig:pk-cubic} and \ref{fig:pk-quartic}. These simulation power spectra were measured using the {\tt POWMES} code \cite{Colombi:2008dw}. We discuss now the performance of the halo model in describing the simulation results, by separating the discussion into large, intermediate and small scales.

{\it Large scales}. On scales $k \lesssim 0.2 h/{\rm Mpc}$, the halo model prediction matches the simulation results. On these scales, the halo model is dominated by the 2-halo term in Eq.~(\ref{eq:halo-model}), which reduces simply to the linear matter power spectrum $P_{k,{\rm lin}}$. More precisely, in the limit in which $k \ll 1 h/{\rm Mpc}$, Eq.~(\ref{eq:I-function}) becomes
\bq\label{eq:I-function-smallk}
I(k) \approx \int {\rm d}M \frac{1}{\bar{\rho}_{m0}}\frac{{\rm d}n(M)}{{\rm dln}M} b_{\rm lin}(M) = 1,
\eq
where we have used the fact that $u(k\rightarrow 0, M) \rightarrow 1$ and the last equality is ensured by the definition of the Sheth-Tormen mass function and linear halo bias (see e.g.~Refs.~\cite{Mo:1996cn, Scoccimarro:2000gm, Cooray:2002dia}). We note that the integral of Eq.~(\ref{eq:I-function-smallk}) is hard to evaluate numerically because, at low $M$, neither the mass function nor the halo bias approach zero. In this paper, we make use of the fact that the last equality of Eq.~(\ref{eq:I-function-smallk}) holds by construction. Effectively, we choose a sufficiently small lower limit ($M \sim 10^6 M_{\odot}/h$), and then simply add the missing contribution to the integral, such that it adds up to unity. We have computed the integrals using both {\tt Python} and {\tt Mathematica} routines, which gave the same results \footnote{The same {\tt Mathematica} routines were used to obtain the results of Ref.~\cite{Lombriser:2013eza}.}. Note also that $P_k^{\rm  2h}$ differs only from $P_{k,{\rm lin}}$ for $k \gtrsim 1 h/{\rm Mpc}$, where $P_k^{\rm 1h}$ already provides the dominant contribution to the total power. This is a general result that is not restricted to the Galileon models studied here \cite{Cooray:2002dia}; in practice, this means that in the halo model approach, it makes almost no difference to use the 2-halo term or the linear matter power spectrum.

{\it Intermediate scales}. On scales $0.2 h/{\rm Mpc} \lesssim k \lesssim 2 h/{\rm Mpc}$, the halo model predictions in all the variants of the Cubic and Quartic Galileon models tend to underpredict the clustering power measured in the simulations. In particular, the mismatch ranges between $\sim 50\%$ and $\sim 20\%$ across all the variants and epochs. These differences are not entirely unexpected and their explanation can be related to some of the approximations associated with the halo model. In particular, the 2-halo term can be written more accurately as
\bq\label{eq:accurate-2halo-term}
P_k^{\rm 2h} = && \int {\rm d}M_1 \frac{1}{\bar{\rho}_{m0}}\frac{{\rm d}n(M_1)}{{\rm dln}M_1} |u(k, M_1)| \nonumber \\
&&\int {\rm d}M_2 \frac{1}{\bar{\rho}_{m0}}\frac{{\rm d}n(M_2)}{{\rm dln}M_2} |u(k, M_2)| P_k^{\rm hh}(M_1, M_2),\nonumber \\
\eq
where $P_k^{\rm hh}(M_1, M_2)$ is the halo-halo power spectrum of haloes with mass $M_1$ and $M_2$. In the standard halo model approach, one approximates $P_k^{\rm hh}(M_1, M_2) = b(M_1)b(M_2)P_{k, {\rm lin}}$, which is done purely for convenience. This way, the two integrals in Eq.~(\ref{eq:accurate-2halo-term}) can be separated and one recovers Eqs.~(\ref{eq:halo-model-terms}) and (\ref{eq:I-function}). This approximation is expected to be valid on large scales. However, on intermediate and small scales, neither the bias parameter nor the matter power spectrum are well approximated by linear theory. Indeed, using the linear halo bias on these scales leads to an overestimation of the power, and using the linear power spectrum leads to an underestimation. As a result, the net effect of these approximations can, in principle, cancel to some extent. Nevertheless, it seems reasonable to expect that this cancellation may not be perfect. Note that this applies not only to the Galileon models studied here, but also to the standard $\Lambda$CDM model. In fact, the {\it halofit} approach is partly motivated as an alternative to the halo model that is more accurate on intermediate scales \cite{Smith:2002dz, Takahashi:2012em}. Recently, Ref.~\cite{Zhao:2013dza} extended the {\it halofit} approach to describe the nonlinear power spectrum in $f(R)$ gravity models. In the lower panels of Figs.~\ref{fig:pk-cubic} and \ref{fig:pk-quartic}, the halo model prediction overestimates the effects of the modifications to gravity, compared to the simulation results. This overestimation is similar to that found in Ref.~\cite{Schmidt:2009yj} for Dvali-Gabadadze-Porrati (DGP) models and Refs.~\cite{Schmidt:2008tn, Lombriser:2013eza} for $f(R)$ models of gravity. However, Ref.~\cite{Lombriser:2013eza} has also shown how a simple modification of the 2-halo term can make the analytical predictions much more accurate.

{\it Small scales}. On scales $k \gtrsim 2 h/{\rm Mpc}$, the agreement between the halo model and simulation results becomes generally better than on intermediate scales, especially at $a = 1$. On these scales, the 1-halo term dominates the total power spectrum, and the good performance of the halo model in matching the power spectrum of the simulations is related to the fact that we have used the fitted mass function and the fitted $c_{200}$($M_{200})$ relation. For instance, in the case of the full variant of the Quartic model at $a = 1$, the use of the standard Sheth-Tormen mass function would significantly overpredict the simulation results (cf.~Fig.~\ref{fig:cmf-quartic}). This would in turn lead to a significant overprediction of the clustering power on small scales as well (not shown to make the plot clearer). Nevertheless, this variant, together with the linear variant of the Cubic model, still shows a visible discrepancy between the halo model and simulation results on these small scales at $a = 1$. In particular, at $k \sim 1 h/{\rm Mpc}$, the halo model of the linear Cubic variant predicts $\sim 30\%$ less power than the simulations; whereas in the case of the full Quartic variant, the halo model overpredicts the power in the simulations by $\sim 40\%$. {Moreover, the performance of the halo model on small scales becomes worse at earlier times. A possible reason for this mismatch can be related to the relaxation state of the haloes. If the haloes are not relaxed, then this can bias the estimation of the concentration, which could explain the differences. To investigate this, we have measured the impact of artifically enhancing and suppressing the amplitude of the concentration-mass relation by 25\% on all mass scales. This test has shown that even a drastic modification of $25\%$ in the halo concentration parameter does not fully reconcile the halo model with the simulations results.  Hence, the discrepancies on small scales are likely to be associated with the approximations of the halo model itself. For instance, recall that the halo model assumes that all the matter in the Universe lies within gravitationally bound structures, which is not the case in N-body simulations. The {\it halofit} approach is also known to be more accurate than the halo model on small scales \cite{Smith:2002dz, Takahashi:2012em}. A more detailed study of the validity of the assumptions of the halo model is beyond the scope of the present paper. In terms of the relative difference, however, these discrepancies cancel to some extent, and the agreement between the halo model and the simulations becomes much better. This is particularly noticeable in the case of the Cubic Galileon model.}

\bq
\nonumber 
\eq

Physically, just as we saw in the previous sections, in Fig.~\ref{fig:pk-cubic}, we reencounter the extreme effectiveness of the screening mechanism in the Cubic model in suppressing any modifications to gravity on small scales. For instance, at $k \sim 1 h/{\rm Mpc}$ and $a = 1$, the increase in power relative to QCDM is of order $10\%$, which is considerably smaller than the $\sim 50\%$ boost seen with the linear variant. The physical picture is much different in Fig.~\ref{fig:pk-quartic} because of the weaker gravity in the Quartic model, which follows from the implementation of the Vainshtein mechanism. In this case, the simulations of the full Quartic model show $\sim 30\%$ less clustering power than QCDM at $k \sim 1 h/{\rm Mpc}$ and $a = 1$; while the simulations of the linear model show an enhancement of about $25\%$. For more details about the results for the power spectrum in the Cubic and Quartic Galileon models we refer the reader to Refs.~\cite{Barreira:2013eea} and \cite{Li:2013tda}, respectively.

\section{Summary}\label{sec:conclusions}

We have studied the properties of dark matter haloes in the Cubic and Quartic Galileon gravity cosmologies. We have made use of N-body simulation results, as well as semi-analytical predictions to investigate the halo mass function, the linear halo bias parameter, the halo concentration-mass relation and the nonlinear matter power spectrum. We have also assessed the performance of the standard semi-analytical formulae in describing the results from the N-body simulations.

The action of these modified gravity models contains derivative self-couplings of a Galilean invariant scalar field, which give rise to a fifth force that depends on the spatial gradients of the scalar field. The magnitude of the fifth force can be made compatible with Solar System tests of gravity by means of the Vainshtein screening mechanism. This dynamically suppresses the spatial variations of the scalar field around massive objects. In the case of the Quartic Galileon model, an additional coupling to the Ricci scalar in the action is necessary to avoid the presence of theoretical instabilities.

We focused on the model parameters that are preferred by the data from the CMB, SNIa and BAO. For these parameter sets, the modifications to gravity are only important at late times and, in low overdensity regions, gravity is enhanced in both models (cf.~Fig.~\ref{fig:Geffmap}). In regions where the density contrast is high, the Vainshtein mechanism effectively suppresses the spatial variations of the Galileon field. This results in a restoration of normal gravity in the Cubic Galileon model, but leads to an overall weakening of the strength of gravity in the Quartic Galileon model. The latter is a consequence of time-varying modifications that follow from the coupling to the Ricci scalar in $\mathcal{L}_4$, which cannot be suppressed by the Vainshtein mechanism. The time-varying nature of the gravitational strength in the Quartic model brings it into severe tension with local gravity constraints. In this paper, however, we focused only on the cosmological consequences of the modifications to gravity. Our main results can be summarized as follows:

\begin{itemize}

\item If one uses the standard Sheth-Tormen fitting parameters $(q,p) = (0.75, 0.30)$, together with the value of $\delta_c$ determined using the spherical collapse model (Table \ref{table:dc-values}), then the formulae for the halo mass function fail to provide a reasonable match to the results of the N-body simulations (see Figs.~\ref{fig:cmf-cubic} and \ref{fig:cmf-quartic}). 

\item By fitting the $(q,p)$ parameters to match the halo mass function measured from the simulations (cf.~Table \ref{table:bf-ap}), then indeed, the Sheth-Tormen formula provides a very good reproduction of the halo abundances over the entire mass range probed by our simulations. {Moreover, the Sheth-Tormen linear halo bias formula computed with the best-fitting $(q,p)$ also provides a good description of the results of the simulations. This shows that the principles of the excursion set theory still hold in Galileon gravity models.}

\item {Previous papers have raised the possibility than the enhanced clustering amplitude of the linear matter power spectrum in the Cubic and Quartic Galileon models ($\sigma_8 \sim 1$, c.f.~Table~\ref{table:table-max}) could potentially lead to some tension with the observed clustering amplitude of LRGs. However, the effect of a boosted linear power spectrum is degenerate with a lower linear halo bias parameter (c.f.~Fig.~\ref{fig:hb}), which can help to ease an eventual tension. In this paper, armed with accurate analytical formulae, we have addressed this issue by performing a halo occupation distribution analysis of LRGs. This has shown that the interplay between the modifications to the large scale clustering power, halo abundance and halo bias in the Cubic and Quartic Galileon models can be explored to yield realistic LRG halo distributions that match both the observed clustering amplitude and galaxy number density. We conclude that the Cubic and Quartic models are unlikely to be in tension with the LRG clustering data.}

\item The halo concentration-mass relation, $c_{200}(M_{200})$, in the Cubic and Quartic Galileon models is well fitted by a power law (cf.~Fig.~\ref{fig:c-m} and Table \ref{table:bf-alphabeta}). The standard picture that the concentration increases with time for fixed mass, and decreases with mass at a given epoch prevails in all but one of the models we studied. The exception is the full variant of the Quartic model, in which the weaker gravity leads to halo concentrations with a very shallow mass and time dependence.

\item On linear scales ($k \lesssim 0.2 h/{\rm Mpc}$), the halo model prediction agrees very well with the matter power spectrum measured from the simulations because both reduce to the linear theory expectation. On intermediate scales ($0.2 h/{\rm Mpc} \lesssim k \lesssim 2 h/{\rm Mpc}$), the halo model typically underpredicts the power spectrum of the simulations by $20\%$ to $50\%$ across all the model variants at all the epochs shown (cf.~Fig.~\ref{fig:pk-cubic} and Fig.~\ref{fig:pk-quartic}). This is a consequence of approximations that are made in the derivation of the halo model equations, which sacrifice accuracy in the mildy-nonlinear regime in favour of analytical convenience. The agreement between the halo model and simulations becomes better on small scales ($k \gtrsim 2 h/{\rm Mpc}$) at $a = 1$. We believe this is closely related to the fact that we have used analytical formulae that match the mass function and concentration parameter of the simulations. There are still visible differences between the formulae and the simulation results on smaller scales for the linear variant of the Cubic and full variant of the Quartic model. Morevoer, these differences also exist for all variants at earlier times. {We have checked that any discrepancies on small scales are likely to be due to the approximations made in the halo model, and not to an incorrect modelling of the halo properties that enter the calculation of the 1-halo term.}

\item In all of our results, we have found that the screening mechanism works very effectively in the Cubic Galileon model, especially on small scales. This is particularly noticeable in our results for the halo concentration-mass relation and the nonlinear matter power spectrum on small scales, for which the full and QCDM variants of the Cubic model give essentially the same predictions. In the case of the Quartic model, the screening mechanism cannot suppress all of the modifications to gravity, which leads to clear differences between the full and QCDM variants of the model. It would be interesting to investigate further the impact of the weaker gravity in this model cosmologically. However, the weaker and time-varying gravitational strength is most likely sufficient to rule this model out on the basis that it does not recover GR in the Solar System. Inevitably, this reduces the interest in pursuing further research in the Quartic Galileon model.

\end{itemize}

With its two free parameters recalibrated, the Sheth-Tormen mass function and its application in the halo model approach, has proven sufficient to give a reasonable match to the results of the Cubic and Quartic Galileon simulations. Although in this paper we focused only on two particular models, we believe that the strategy presented here of directly fitting the halo properties to simulations can also be applied to other modified gravity theories \cite{Gaztanaga:2000vw, Schaefer:2007nf, Martino:2008ae, Li:2011qda, Borisov:2011fu, Lombriser:2013wta, Kopp:2013lea, Taddei:2013bsk}. For some models, we believe this can improve the performance of the semi-analytical formulae. These are a much faster alternative to N-body simulations, and can be used to generate quick estimates for the large-scale structure in modified gravity. The development of these semi-analytical models, in addition to enabling a clearer way of pinpointing the physical effects, can also be of crucial importance for current and upcoming large-scale structure surveys, which will require vast regions of parameter spaces to be spanned in a timely manner.

\begin{acknowledgments}

We thank Yan-Chuan Cai for useful comments and discussions. This work used the DiRAC Data Centric system at Durham University, operated by the Institute for Computational Cosmology on behalf of the STFC DiRAC HPC Facility (\url{www.dirac.ac.uk}). This equipment was funded by BIS National E-infrastructure capital grant ST/K00042X/1, STFC capital grant ST/H008519/1, and STFC DiRAC Operations grant ST/K003267/1 and Durham University. DiRAC is part of the National E-Infrastructure. AB is supported by FCT-Portugal through grant SFRH/BD/75791/2011. BL is supported by the Royal Astronomical Society and Durham University. WAH is supported by the Polish National Science Center through grant DEC-2011/01/D/ST9/01960 and ERC Advanced Investigator grant of C. S. Frenk, COSMIWAY. LL is supported by the STFC Consolidated Grant for Astronomy and Astrophysics at the University of Edinburgh. This work has been partially supported by the European Union FP7  ITN INVISIBLES (Marie Curie Actions, PITN- GA-2011- 289442) and STFC. 

\end{acknowledgments}

\bibliography{halo-model-galileon.bib}

\begin{thebibliography}{10}%
\makeatletter
\providecommand \@ifxundefined [1]{%
 \ifx #1\undefined \expandafter \@firstoftwo
 \else \expandafter \@secondoftwo
\fi
}%
\providecommand \@ifnum [1]{%
 \ifnum #1\expandafter \@firstoftwo
 \else \expandafter \@secondoftwo
\fi
}%
\providecommand \enquote [1]{``#1''}%
\providecommand \bibnamefont  [1]{#1}%
\providecommand \bibfnamefont [1]{#1}%
\providecommand \citenamefont [1]{#1}%
\providecommand\href[0]{\@sanitize\@href}%
\providecommand\@href[1]{\endgroup\@@startlink{#1}\endgroup\@@href}%
\providecommand\@@href[1]{#1\@@endlink}%
\providecommand \@sanitize [0]{\begingroup\catcode`\&12\catcode`\#12\relax}%
\@ifxundefined \pdfoutput {\@firstoftwo}{%
 \@ifnum{\z@=\pdfoutput}{\@firstoftwo}{\@secondoftwo}%
}{%
 \providecommand\@@startlink[1]{\leavevmode\special{html:<a href="#1">}}%
 \providecommand\@@endlink[0]{\special{html:</a>}}%
}{%
 \providecommand\@@startlink[1]{%
  \leavevmode
  \pdfstartlink
   attr{/Border[0 0 1 ]/H/I/C[0 1 1]}%
   user{/Subtype/Link/A<</Type/Action/S/URI/URI(#1)>>}%
  \relax
 }%
 \providecommand\@@endlink[0]{\pdfendlink}%
}%
\providecommand \url  [0]{\begingroup\@sanitize \@url }%
\providecommand \@url [1]{\endgroup\@href {#1}{\urlprefix}}%
\providecommand \urlprefix [0]{URL }%
\providecommand \Eprint[0]{\href }%
\@ifxundefined \urlstyle {%
  \providecommand \doi [1]{doi:\discretionary{}{}{}#1}%
}{%
  \providecommand \doi [0]{doi:\discretionary{}{}{}\begingroup
  \urlstyle{rm}\Url }%
}%
\providecommand \doibase [0]{http://dx.doi.org/}%
\providecommand \Doi[1]{\href{\doibase#1}}%
\providecommand \bibAnnote [3]{%
  \BibitemShut{#1}%
  \begin{quotation}\noindent
    \textsc{Key:}\ #2\\\textsc{Annotation:}\ #3%
  \end{quotation}%
}%
\providecommand \bibAnnoteFile [2]{%
  \IfFileExists{#2}{\bibAnnote {#1} {#2} {\input{#2}}}{}%
}%
\providecommand \typeout [0]{\immediate \write \m@ne }%
\providecommand \selectlanguage [0]{\@gobble}%
\providecommand \bibinfo [0]{\@secondoftwo}%
\providecommand \bibfield [0]{\@secondoftwo}%
\providecommand \translation [1]{[#1]}%
\providecommand \BibitemOpen[0]{}%
\providecommand \bibitemStop [0]{}%
\providecommand \bibitemNoStop [0]{.\EOS\space}%
\providecommand \EOS [0]{\spacefactor3000\relax}%
\providecommand \BibitemShut [1]{\csname bibitem#1\endcsname}%
\bibitem{Clifton:2011jh}%
  \BibitemOpen
  \bibfield{author}{%
  \bibinfo {author} {\bibfnamefont{T.}~\bibnamefont{Clifton}}, \bibinfo
  {author} {\bibfnamefont{P.~G.}\ \bibnamefont{Ferreira}}, \bibinfo {author}
  {\bibfnamefont{A.}~\bibnamefont{Padilla}},\ and\ \bibinfo {author}
  {\bibfnamefont{C.}~\bibnamefont{Skordis}},\ }%
  \bibfield{journal}{%
  \Doi{10.1016/j.physrep.2012.01.001}{\bibinfo {journal} {Phys.Rept.}}\ }%
  \textbf{\bibinfo {volume} {513}},\ \bibinfo {pages} {1} (\bibinfo {year}
  {2012}),\ \Eprint{http://arxiv.org/abs/1106.2476}{arXiv:1106.2476
  [astro-ph.CO]}%
  \bibAnnoteFile{NoStop}{Clifton:2011jh}%
\bibitem{PhysRevD.79.064036}%
  \BibitemOpen
  \bibfield{author}{%
  \bibinfo {author} {\bibfnamefont{A.}~\bibnamefont{Nicolis}}, \bibinfo
  {author} {\bibfnamefont{R.}~\bibnamefont{Rattazzi}},\ and\ \bibinfo {author}
  {\bibfnamefont{E.}~\bibnamefont{Trincherini}},\ }%
  \bibfield{journal}{%
  \Doi{10.1103/PhysRevD.79.064036}{\bibinfo {journal} {Phys. Rev. D}}\ }%
  \textbf{\bibinfo {volume} {79}},\ \bibinfo {pages} {064036} (\bibinfo {year}
  {2009})%
  \bibAnnoteFile{NoStop}{PhysRevD.79.064036}%
\bibitem{Woodard:2006nt}%
  \BibitemOpen
  \bibfield{author}{%
  \bibinfo {author} {\bibfnamefont{R.~P.}\ \bibnamefont{Woodard}},\ }%
  \bibfield{journal}{%
  \Doi{10.1007/978-3-540-71013-4_14}{\bibinfo {journal} {Lect.Notes Phys.}}\ }%
  \textbf{\bibinfo {volume} {720}},\ \bibinfo {pages} {403} (\bibinfo {year}
  {2007}),\
  \Eprint{http://arxiv.org/abs/astro-ph/0601672}{arXiv:astro-ph/0601672
  [astro-ph]}%
  \bibAnnoteFile{NoStop}{Woodard:2006nt}%
\bibitem{PhysRevD.79.084003}%
  \BibitemOpen
  \bibfield{author}{%
  \bibinfo {author} {\bibfnamefont{C.}~\bibnamefont{Deffayet}}, \bibinfo
  {author} {\bibfnamefont{G.}~\bibnamefont{Esposito-Far\`ese}},\ and\ \bibinfo
  {author} {\bibfnamefont{A.}~\bibnamefont{Vikman}},\ }%
  \bibfield{journal}{%
  \Doi{10.1103/PhysRevD.79.084003}{\bibinfo {journal} {Phys. Rev. D}}\ }%
  \textbf{\bibinfo {volume} {79}},\ \bibinfo {pages} {084003} (\bibinfo {year}
  {2009})%
  \bibAnnoteFile{NoStop}{PhysRevD.79.084003}%
\bibitem{Deffayet:2009mn}%
  \BibitemOpen
  \bibfield{author}{%
  \bibinfo {author} {\bibfnamefont{C.}~\bibnamefont{Deffayet}}, \bibinfo
  {author} {\bibfnamefont{S.}~\bibnamefont{Deser}},\ and\ \bibinfo {author}
  {\bibfnamefont{G.}~\bibnamefont{Esposito-Farese}},\ }%
  \bibfield{journal}{%
  \Doi{10.1103/PhysRevD.80.064015}{\bibinfo {journal} {Phys.Rev.}}\ }%
  \textbf{\bibinfo {volume} {D80}},\ \bibinfo {pages} {064015} (\bibinfo {year}
  {2009}),\ \Eprint{http://arxiv.org/abs/0906.1967}{arXiv:0906.1967 [gr-qc]}%
  \bibAnnoteFile{NoStop}{Deffayet:2009mn}%
\bibitem{Horndeski:1974wa}%
  \BibitemOpen
  \bibfield{author}{%
  \bibinfo {author} {\bibfnamefont{G.~W.}\ \bibnamefont{Horndeski}},\ }%
  \bibfield{journal}{%
  \Doi{10.1007/BF01807638}{\bibinfo {journal} {Int.J.Theor.Phys.}}\ }%
  \textbf{\bibinfo {volume} {10}},\ \bibinfo {pages} {363} (\bibinfo {year}
  {1974})%
  \bibAnnoteFile{NoStop}{Horndeski:1974wa}%
\bibitem{Will:2005va}%
  \BibitemOpen
  \bibfield{author}{%
  \bibinfo {author} {\bibfnamefont{C.~M.}\ \bibnamefont{Will}},\ }%
  \bibfield{journal}{%
  \bibinfo {journal} {Living Rev.Rel.}\ }%
  \textbf{\bibinfo {volume} {9}},\ \bibinfo {pages} {3} (\bibinfo {year}
  {2006}),\ \Eprint{http://arxiv.org/abs/gr-qc/0510072}{arXiv:gr-qc/0510072
  [gr-qc]}%
  \bibAnnoteFile{NoStop}{Will:2005va}%
\bibitem{Vainshtein1972393}%
  \BibitemOpen
  \bibfield{author}{%
  \bibinfo {author} {\bibfnamefont{A.}~\bibnamefont{Vainshtein}},\ }%
  \bibfield{journal}{%
  \Doi{10.1016/0370-2693(72)90147-5}{\bibinfo {journal} {Phys.~Lett.~B}}\ }%
  \textbf{\bibinfo {volume} {39}},\ \bibinfo {pages} {393 } (\bibinfo {year}
  {1972}),\ ISSN \bibinfo {issn} {0370-2693}%
  \bibAnnoteFile{NoStop}{Vainshtein1972393}%
\bibitem{Babichev:2013usa}%
  \BibitemOpen
  \bibfield{author}{%
  \bibinfo {author} {\bibfnamefont{E.}~\bibnamefont{Babichev}}\ and\ \bibinfo
  {author} {\bibfnamefont{C.}~\bibnamefont{Deffayet}}}%
   (\bibinfo {year} {2013}),\
  \Eprint{http://arxiv.org/abs/1304.7240}{arXiv:1304.7240 [gr-qc]}%
  \bibAnnoteFile{NoStop}{Babichev:2013usa}%
\bibitem{Koyama:2013paa}%
  \BibitemOpen
  \bibfield{author}{%
  \bibinfo {author} {\bibfnamefont{K.}~\bibnamefont{Koyama}}, \bibinfo {author}
  {\bibfnamefont{G.}~\bibnamefont{Niz}},\ and\ \bibinfo {author}
  {\bibfnamefont{G.}~\bibnamefont{Tasinato}},\ }%
  \bibfield{journal}{%
  \Doi{10.1103/PhysRevD.88.021502}{\bibinfo {journal} {Phys.Rev.}}\ }%
  \textbf{\bibinfo {volume} {D88}},\ \bibinfo {pages} {021502} (\bibinfo {year}
  {2013}),\ \Eprint{http://arxiv.org/abs/1305.0279}{arXiv:1305.0279 [hep-th]}%
  \bibAnnoteFile{NoStop}{Koyama:2013paa}%
\bibitem{Barreira:2012kk}%
  \BibitemOpen
  \bibfield{author}{%
  \bibinfo {author} {\bibfnamefont{A.}~\bibnamefont{Barreira}}, \bibinfo
  {author} {\bibfnamefont{B.}~\bibnamefont{Li}}, \bibinfo {author}
  {\bibfnamefont{C.~M.}\ \bibnamefont{Baugh}},\ and\ \bibinfo {author}
  {\bibfnamefont{S.}~\bibnamefont{Pascoli}},\ }%
  \bibfield{journal}{%
  \Doi{10.1103/PhysRevD.86.124016}{\bibinfo {journal} {Phys.Rev.}}\ }%
  \textbf{\bibinfo {volume} {D86}},\ \bibinfo {pages} {124016} (\bibinfo {year}
  {2012}),\ \Eprint{http://arxiv.org/abs/1208.0600}{arXiv:1208.0600
  [astro-ph.CO]}%
  \bibAnnoteFile{NoStop}{Barreira:2012kk}%
\bibitem{Barreira:2013jma}%
  \BibitemOpen
  \bibfield{author}{%
  \bibinfo {author} {\bibfnamefont{A.}~\bibnamefont{Barreira}}, \bibinfo
  {author} {\bibfnamefont{B.}~\bibnamefont{Li}}, \bibinfo {author}
  {\bibfnamefont{A.}~\bibnamefont{Sanchez}}, \bibinfo {author}
  {\bibfnamefont{C.~M.}\ \bibnamefont{Baugh}},\ and\ \bibinfo {author}
  {\bibfnamefont{S.}~\bibnamefont{Pascoli}},\ }%
  \bibfield{journal}{%
  \Doi{10.1103/PhysRevD.87.103511}{\bibinfo {journal} {Phys.Rev.}}\ }%
  \textbf{\bibinfo {volume} {D87}},\ \bibinfo {pages} {103511} (\bibinfo {year}
  {2013}),\ \Eprint{http://arxiv.org/abs/1302.6241}{arXiv:1302.6241
  [astro-ph.CO]}%
  \bibAnnoteFile{NoStop}{Barreira:2013jma}%
\bibitem{Gannouji:2010au}%
  \BibitemOpen
  \bibfield{author}{%
  \bibinfo {author} {\bibfnamefont{R.}~\bibnamefont{Gannouji}}\ and\ \bibinfo
  {author} {\bibfnamefont{M.}~\bibnamefont{Sami}},\ }%
  \bibfield{journal}{%
  \Doi{10.1103/PhysRevD.82.024011}{\bibinfo {journal} {Phys.Rev.}}\ }%
  \textbf{\bibinfo {volume} {D82}},\ \bibinfo {pages} {024011} (\bibinfo {year}
  {2010}),\ \Eprint{http://arxiv.org/abs/1004.2808}{arXiv:1004.2808 [gr-qc]}%
  \bibAnnoteFile{NoStop}{Gannouji:2010au}%
\bibitem{PhysRevD.80.024037}%
  \BibitemOpen
  \bibfield{author}{%
  \bibinfo {author} {\bibfnamefont{N.}~\bibnamefont{Chow}}\ and\ \bibinfo
  {author} {\bibfnamefont{J.}~\bibnamefont{Khoury}},\ }%
  \bibfield{journal}{%
  \Doi{10.1103/PhysRevD.80.024037}{\bibinfo {journal} {Phys. Rev. D}}\ }%
  \textbf{\bibinfo {volume} {80}},\ \bibinfo {pages} {024037} (\bibinfo {year}
  {2009})%
  \bibAnnoteFile{NoStop}{PhysRevD.80.024037}%
\bibitem{DeFelice:2010pv}%
  \BibitemOpen
  \bibfield{author}{%
  \bibinfo {author} {\bibfnamefont{A.}~\bibnamefont{De~Felice}}\ and\ \bibinfo
  {author} {\bibfnamefont{S.}~\bibnamefont{Tsujikawa}},\ }%
  \bibfield{journal}{%
  \Doi{10.1103/PhysRevLett.105.111301}{\bibinfo {journal} {Phys.Rev.Lett.}}\ }%
  \textbf{\bibinfo {volume} {105}},\ \bibinfo {pages} {111301} (\bibinfo {year}
  {2010}),\ \Eprint{http://arxiv.org/abs/1007.2700}{arXiv:1007.2700
  [astro-ph.CO]}%
  \bibAnnoteFile{NoStop}{DeFelice:2010pv}%
\bibitem{Nesseris:2010pc}%
  \BibitemOpen
  \bibfield{author}{%
  \bibinfo {author} {\bibfnamefont{S.}~\bibnamefont{Nesseris}}, \bibinfo
  {author} {\bibfnamefont{A.}~\bibnamefont{De~Felice}},\ and\ \bibinfo {author}
  {\bibfnamefont{S.}~\bibnamefont{Tsujikawa}},\ }%
  \bibfield{journal}{%
  \Doi{10.1103/PhysRevD.82.124054}{\bibinfo {journal} {Phys.Rev.}}\ }%
  \textbf{\bibinfo {volume} {D82}},\ \bibinfo {pages} {124054} (\bibinfo {year}
  {2010}),\ \Eprint{http://arxiv.org/abs/1010.0407}{arXiv:1010.0407
  [astro-ph.CO]}%
  \bibAnnoteFile{NoStop}{Nesseris:2010pc}%
\bibitem{Appleby:2011aa}%
  \BibitemOpen
  \bibfield{author}{%
  \bibinfo {author} {\bibfnamefont{S.~A.}\ \bibnamefont{Appleby}}\ and\
  \bibinfo {author} {\bibfnamefont{E.~V.}\ \bibnamefont{Linder}},\ }%
  \bibfield{journal}{%
  \bibinfo {journal} {JCAP}\ }%
  \textbf{\bibinfo {volume} {1203}},\ \bibinfo {pages} {043} (\bibinfo {year}
  {2012}),\ \Eprint{http://arxiv.org/abs/1112.1981}{arXiv:1112.1981
  [astro-ph.CO]}%
  \bibAnnoteFile{NoStop}{Appleby:2011aa}%
\bibitem{PhysRevD.82.103015}%
  \BibitemOpen
  \bibfield{author}{%
  \bibinfo {author} {\bibfnamefont{A.}~\bibnamefont{Ali}}, \bibinfo {author}
  {\bibfnamefont{R.}~\bibnamefont{Gannouji}},\ and\ \bibinfo {author}
  {\bibfnamefont{M.}~\bibnamefont{Sami}},\ }%
  \bibfield{journal}{%
  \Doi{10.1103/PhysRevD.82.103015}{\bibinfo {journal} {Phys. Rev. D}}\ }%
  \textbf{\bibinfo {volume} {82}},\ \bibinfo {pages} {103015} (\bibinfo {year}
  {2010})%
  \bibAnnoteFile{NoStop}{PhysRevD.82.103015}%
\bibitem{Neveu:2013mfa}%
  \BibitemOpen
  \bibfield{author}{%
  \bibinfo {author} {\bibfnamefont{J.}~\bibnamefont{Neveu}}, \bibinfo {author}
  {\bibfnamefont{V.}~\bibnamefont{Ruhlmann-Kleider}}, \bibinfo {author}
  {\bibfnamefont{A.}~\bibnamefont{Conley}}, \bibinfo {author}
  {\bibfnamefont{N.}~\bibnamefont{Palanque-Delabrouille}}, \bibinfo {author}
  {\bibfnamefont{P.}~\bibnamefont{Astier}}, \emph{et~al.}}%
   (\bibinfo {year} {2013}),\
  \Eprint{http://arxiv.org/abs/1302.2786}{arXiv:1302.2786 [gr-qc]}%
  \bibAnnoteFile{NoStop}{Neveu:2013mfa}%
\bibitem{Appleby:2012ba}%
  \BibitemOpen
  \bibfield{author}{%
  \bibinfo {author} {\bibfnamefont{S.~A.}\ \bibnamefont{Appleby}}\ and\
  \bibinfo {author} {\bibfnamefont{E.~V.}\ \bibnamefont{Linder}}}%
   (\bibinfo {year} {2012}),\
  \Eprint{http://arxiv.org/abs/1204.4314}{arXiv:1204.4314 [astro-ph.CO]}%
  \bibAnnoteFile{NoStop}{Appleby:2012ba}%
\bibitem{Okada:2012mn}%
  \BibitemOpen
  \bibfield{author}{%
  \bibinfo {author} {\bibfnamefont{H.}~\bibnamefont{Okada}}, \bibinfo {author}
  {\bibfnamefont{T.}~\bibnamefont{Totani}},\ and\ \bibinfo {author}
  {\bibfnamefont{S.}~\bibnamefont{Tsujikawa}}}%
   (\bibinfo {year} {2012}),\
  \Eprint{http://arxiv.org/abs/1208.4681}{arXiv:1208.4681 [astro-ph.CO]}%
  \bibAnnoteFile{NoStop}{Okada:2012mn}%
\bibitem{Bartolo:2013ws}%
  \BibitemOpen
  \bibfield{author}{%
  \bibinfo {author} {\bibfnamefont{N.}~\bibnamefont{Bartolo}}, \bibinfo
  {author} {\bibfnamefont{E.}~\bibnamefont{Bellini}}, \bibinfo {author}
  {\bibfnamefont{D.}~\bibnamefont{Bertacca}},\ and\ \bibinfo {author}
  {\bibfnamefont{S.}~\bibnamefont{Matarrese}}}%
   (\bibinfo {year} {2013}),\
  \Eprint{http://arxiv.org/abs/1301.4831}{arXiv:1301.4831 [astro-ph.CO]}%
  \bibAnnoteFile{NoStop}{Bartolo:2013ws}%
\bibitem{Hinshaw:2012fq}%
  \BibitemOpen
  \bibfield{author}{%
  \bibinfo {author} {\bibfnamefont{G.}~\bibnamefont{Hinshaw}}, \bibinfo
  {author} {\bibfnamefont{D.}~\bibnamefont{Larson}}, \bibinfo {author}
  {\bibfnamefont{E.}~\bibnamefont{Komatsu}}, \bibinfo {author}
  {\bibfnamefont{D.}~\bibnamefont{Spergel}}, \bibinfo {author}
  {\bibfnamefont{C.}~\bibnamefont{Bennett}}, \emph{et~al.}}%
   (\bibinfo {year} {2012}),\
  \Eprint{http://arxiv.org/abs/1212.5226}{arXiv:1212.5226 [astro-ph.CO]}%
  \bibAnnoteFile{NoStop}{Hinshaw:2012fq}%
\bibitem{Ade:2013zuv}%
  \BibitemOpen
  \bibfield{author}{%
  \bibinfo {author} {\bibfnamefont{P.}~\bibnamefont{Ade}} \emph{et~al.}
  (\bibinfo {collaboration} {Planck Collaboration})}%
   (\bibinfo {year} {2013}),\
  \Eprint{http://arxiv.org/abs/1303.5076}{arXiv:1303.5076 [astro-ph.CO]}%
  \bibAnnoteFile{NoStop}{Ade:2013zuv}%
\bibitem{Reid:2009xm}%
  \BibitemOpen
  \bibfield{author}{%
  \bibinfo {author} {\bibfnamefont{B.~A.}\ \bibnamefont{Reid}}, \bibinfo
  {author} {\bibfnamefont{W.~J.}\ \bibnamefont{Percival}}, \bibinfo {author}
  {\bibfnamefont{D.~J.}\ \bibnamefont{Eisenstein}}, \bibinfo {author}
  {\bibfnamefont{L.}~\bibnamefont{Verde}}, \bibinfo {author}
  {\bibfnamefont{D.~N.}\ \bibnamefont{Spergel}}, \emph{et~al.},\ }%
  \bibfield{journal}{%
  \Doi{10.1111/j.1365-2966.2010.16276.x}{\bibinfo {journal}
  {Mon.Not.Roy.Astron.Soc.}}\ }%
  \textbf{\bibinfo {volume} {404}},\ \bibinfo {pages} {60} (\bibinfo {year}
  {2010}),\ \Eprint{http://arxiv.org/abs/0907.1659}{arXiv:0907.1659
  [astro-ph.CO]}%
  \bibAnnoteFile{NoStop}{Reid:2009xm}%
\bibitem{Barreira:2013eea}%
  \BibitemOpen
  \bibfield{author}{%
  \bibinfo {author} {\bibfnamefont{A.}~\bibnamefont{Barreira}}, \bibinfo
  {author} {\bibfnamefont{B.}~\bibnamefont{Li}}, \bibinfo {author}
  {\bibfnamefont{W.~A.}\ \bibnamefont{Hellwing}}, \bibinfo {author}
  {\bibfnamefont{C.~M.}\ \bibnamefont{Baugh}},\ and\ \bibinfo {author}
  {\bibfnamefont{S.}~\bibnamefont{Pascoli}},\ }%
  \bibfield{journal}{%
  \bibinfo {journal} {JCAP}\ }%
  \textbf{\bibinfo {volume} {2013}},\ \bibinfo {pages} {027} (\bibinfo {year}
  {2013}),\ \Eprint{http://arxiv.org/abs/1306.3219}{arXiv:1306.3219
  [astro-ph.CO]}%
  \bibAnnoteFile{NoStop}{Barreira:2013eea}%
\bibitem{Li:2013tda}%
  \BibitemOpen
  \bibfield{author}{%
  \bibinfo {author} {\bibfnamefont{B.}~\bibnamefont{Li}}, \bibinfo {author}
  {\bibfnamefont{A.}~\bibnamefont{Barreira}}, \bibinfo {author}
  {\bibfnamefont{C.~M.}\ \bibnamefont{Baugh}}, \bibinfo {author}
  {\bibfnamefont{W.~A.}\ \bibnamefont{Hellwing}}, \bibinfo {author}
  {\bibfnamefont{K.}~\bibnamefont{Koyama}}, \emph{et~al.}\ }%
  \textbf{\bibinfo {volume} {JCAP11}},\ \bibinfo {pages} {012} (\bibinfo {year}
  {2013}),\ \Eprint{http://arxiv.org/abs/1308.3491}{arXiv:1308.3491
  [astro-ph.CO]}%
  \bibAnnoteFile{NoStop}{Li:2013tda}%
\bibitem{Barreira:2013xea}%
  \BibitemOpen
  \bibfield{author}{%
  \bibinfo {author} {\bibfnamefont{A.}~\bibnamefont{Barreira}}, \bibinfo
  {author} {\bibfnamefont{B.}~\bibnamefont{Li}}, \bibinfo {author}
  {\bibfnamefont{C.}~\bibnamefont{Baugh}},\ and\ \bibinfo {author}
  {\bibfnamefont{S.}~\bibnamefont{Pascoli}}}%
   (\bibinfo {year} {2013}),\
  \Eprint{http://arxiv.org/abs/1308.3699}{arXiv:1308.3699 [astro-ph.CO]}%
  \bibAnnoteFile{NoStop}{Barreira:2013xea}%
\bibitem{Sheth:1999mn}%
  \BibitemOpen
  \bibfield{author}{%
  \bibinfo {author} {\bibfnamefont{R.~K.}\ \bibnamefont{Sheth}}\ and\ \bibinfo
  {author} {\bibfnamefont{G.}~\bibnamefont{Tormen}},\ }%
  \bibfield{journal}{%
  \Doi{10.1046/j.1365-8711.1999.02692.x}{\bibinfo {journal}
  {Mon.Not.Roy.Astron.Soc.}}\ }%
  \textbf{\bibinfo {volume} {308}},\ \bibinfo {pages} {119} (\bibinfo {year}
  {1999}),\
  \Eprint{http://arxiv.org/abs/astro-ph/9901122}{arXiv:astro-ph/9901122
  [astro-ph]}%
  \bibAnnoteFile{NoStop}{Sheth:1999mn}%
\bibitem{Sheth:1999su}%
  \BibitemOpen
  \bibfield{author}{%
  \bibinfo {author} {\bibfnamefont{R.~K.}\ \bibnamefont{Sheth}}, \bibinfo
  {author} {\bibfnamefont{H.}~\bibnamefont{Mo}},\ and\ \bibinfo {author}
  {\bibfnamefont{G.}~\bibnamefont{Tormen}},\ }%
  \bibfield{journal}{%
  \Doi{10.1046/j.1365-8711.2001.04006.x}{\bibinfo {journal}
  {Mon.Not.Roy.Astron.Soc.}}\ }%
  \textbf{\bibinfo {volume} {323}},\ \bibinfo {pages} {1} (\bibinfo {year}
  {2001}),\
  \Eprint{http://arxiv.org/abs/astro-ph/9907024}{arXiv:astro-ph/9907024
  [astro-ph]}%
  \bibAnnoteFile{NoStop}{Sheth:1999su}%
\bibitem{Sheth:2001dp}%
  \BibitemOpen
  \bibfield{author}{%
  \bibinfo {author} {\bibfnamefont{R.~K.}\ \bibnamefont{Sheth}}\ and\ \bibinfo
  {author} {\bibfnamefont{G.}~\bibnamefont{Tormen}},\ }%
  \bibfield{journal}{%
  \Doi{10.1046/j.1365-8711.2002.04950.x}{\bibinfo {journal}
  {Mon.Not.Roy.Astron.Soc.}}\ }%
  \textbf{\bibinfo {volume} {329}},\ \bibinfo {pages} {61} (\bibinfo {year}
  {2002}),\
  \Eprint{http://arxiv.org/abs/astro-ph/0105113}{arXiv:astro-ph/0105113
  [astro-ph]}%
  \bibAnnoteFile{NoStop}{Sheth:2001dp}%
\bibitem{Cooray:2002dia}%
  \BibitemOpen
  \bibfield{author}{%
  \bibinfo {author} {\bibfnamefont{A.}~\bibnamefont{Cooray}}\ and\ \bibinfo
  {author} {\bibfnamefont{R.~K.}\ \bibnamefont{Sheth}},\ }%
  \bibfield{journal}{%
  \Doi{10.1016/S0370-1573(02)00276-4}{\bibinfo {journal} {Phys.Rept.}}\ }%
  \textbf{\bibinfo {volume} {372}},\ \bibinfo {pages} {1} (\bibinfo {year}
  {2002}),\
  \Eprint{http://arxiv.org/abs/astro-ph/0206508}{arXiv:astro-ph/0206508
  [astro-ph]}%
  \bibAnnoteFile{NoStop}{Cooray:2002dia}%
\bibitem{Navarro:1996gj}%
  \BibitemOpen
  \bibfield{author}{%
  \bibinfo {author} {\bibfnamefont{J.~F.}\ \bibnamefont{Navarro}}, \bibinfo
  {author} {\bibfnamefont{C.~S.}\ \bibnamefont{Frenk}},\ and\ \bibinfo {author}
  {\bibfnamefont{S.~D.}\ \bibnamefont{White}},\ }%
  \bibfield{journal}{%
  \Doi{10.1086/304888}{\bibinfo {journal} {Astrophys.J.}}\ }%
  \textbf{\bibinfo {volume} {490}},\ \bibinfo {pages} {493} (\bibinfo {year}
  {1997}),\
  \Eprint{http://arxiv.org/abs/astro-ph/9611107}{arXiv:astro-ph/9611107
  [astro-ph]}%
  \bibAnnoteFile{NoStop}{Navarro:1996gj}%
\bibitem{Deffayet:2010qz}%
  \BibitemOpen
  \bibfield{author}{%
  \bibinfo {author} {\bibfnamefont{C.}~\bibnamefont{Deffayet}}, \bibinfo
  {author} {\bibfnamefont{O.}~\bibnamefont{Pujolas}}, \bibinfo {author}
  {\bibfnamefont{I.}~\bibnamefont{Sawicki}},\ and\ \bibinfo {author}
  {\bibfnamefont{A.}~\bibnamefont{Vikman}},\ }%
  \bibfield{journal}{%
  \Doi{10.1088/1475-7516/2010/10/026}{\bibinfo {journal} {JCAP}}\ }%
  \textbf{\bibinfo {volume} {1010}},\ \bibinfo {pages} {026} (\bibinfo {year}
  {2010}),\ \Eprint{http://arxiv.org/abs/1008.0048}{arXiv:1008.0048 [hep-th]}%
  \bibAnnoteFile{NoStop}{Deffayet:2010qz}%
\bibitem{Babichev:2012re}%
  \BibitemOpen
  \bibfield{author}{%
  \bibinfo {author} {\bibfnamefont{E.}~\bibnamefont{Babichev}}\ and\ \bibinfo
  {author} {\bibfnamefont{G.}~\bibnamefont{Esposito-Farèse}},\ }%
  \bibfield{journal}{%
  \Doi{10.1103/PhysRevD.87.044032}{\bibinfo {journal} {Phys.Rev.}}\ }%
  \textbf{\bibinfo {volume} {D87}},\ \bibinfo {pages} {044032} (\bibinfo {year}
  {2013}),\ \Eprint{http://arxiv.org/abs/1212.1394}{arXiv:1212.1394 [gr-qc]}%
  \bibAnnoteFile{NoStop}{Babichev:2012re}%
\bibitem{Kimura:2010di}%
  \BibitemOpen
  \bibfield{author}{%
  \bibinfo {author} {\bibfnamefont{R.}~\bibnamefont{Kimura}}\ and\ \bibinfo
  {author} {\bibfnamefont{K.}~\bibnamefont{Yamamoto}},\ }%
  \bibfield{journal}{%
  \Doi{10.1088/1475-7516/2011/04/025}{\bibinfo {journal} {JCAP}}\ }%
  \textbf{\bibinfo {volume} {1104}},\ \bibinfo {pages} {025} (\bibinfo {year}
  {2011}),\ \Eprint{http://arxiv.org/abs/1011.2006}{arXiv:1011.2006
  [astro-ph.CO]}%
  \bibAnnoteFile{NoStop}{Kimura:2010di}%
\bibitem{Guy:2010bc}%
  \BibitemOpen
  \bibfield{author}{%
  \bibinfo {author} {\bibfnamefont{J.}~\bibnamefont{Guy}}, \bibinfo {author}
  {\bibfnamefont{M.}~\bibnamefont{Sullivan}}, \bibinfo {author}
  {\bibfnamefont{A.}~\bibnamefont{Conley}}, \bibinfo {author}
  {\bibfnamefont{N.}~\bibnamefont{Regnault}}, \bibinfo {author}
  {\bibfnamefont{P.}~\bibnamefont{Astier}}, \emph{et~al.},\ }%
  \bibfield{journal}{%
  \Doi{10.1051/0004-6361/201014468}{\bibinfo {journal} {Astron.Astrophys.}}\ }%
  \textbf{\bibinfo {volume} {523}},\ \bibinfo {pages} {A7} (\bibinfo {year}
  {2010}),\ \Eprint{http://arxiv.org/abs/1010.4743}{arXiv:1010.4743
  [astro-ph.CO]}%
  \bibAnnoteFile{NoStop}{Guy:2010bc}%
\bibitem{Beutler:2011hx}%
  \BibitemOpen
  \bibfield{author}{%
  \bibinfo {author} {\bibfnamefont{F.}~\bibnamefont{Beutler}}, \bibinfo
  {author} {\bibfnamefont{C.}~\bibnamefont{Blake}}, \bibinfo {author}
  {\bibfnamefont{M.}~\bibnamefont{Colless}}, \bibinfo {author}
  {\bibfnamefont{D.~H.}\ \bibnamefont{Jones}}, \bibinfo {author}
  {\bibfnamefont{L.}~\bibnamefont{Staveley-Smith}}, \emph{et~al.},\ }%
  \bibfield{journal}{%
  \Doi{10.1111/j.1365-2966.2011.19250.x}{\bibinfo {journal}
  {Mon.Not.Roy.Astron.Soc.}}\ }%
  \textbf{\bibinfo {volume} {416}},\ \bibinfo {pages} {3017} (\bibinfo {year}
  {2011}),\ \Eprint{http://arxiv.org/abs/1106.3366}{arXiv:1106.3366
  [astro-ph.CO]}%
  \bibAnnoteFile{NoStop}{Beutler:2011hx}%
\bibitem{Percival:2009xn}%
  \BibitemOpen
  \bibfield{author}{%
  \bibinfo {author} {\bibfnamefont{W.~J.}\ \bibnamefont{Percival}}
  \emph{et~al.} (\bibinfo {collaboration} {SDSS Collaboration}),\ }%
  \bibfield{journal}{%
  \Doi{10.1111/j.1365-2966.2009.15812.x}{\bibinfo {journal}
  {Mon.Not.Roy.Astron.Soc.}}\ }%
  \textbf{\bibinfo {volume} {401}},\ \bibinfo {pages} {2148} (\bibinfo {year}
  {2010}),\ \Eprint{http://arxiv.org/abs/0907.1660}{arXiv:0907.1660
  [astro-ph.CO]}%
  \bibAnnoteFile{NoStop}{Percival:2009xn}%
\bibitem{Anderson:2012sa}%
  \BibitemOpen
  \bibfield{author}{%
  \bibinfo {author} {\bibfnamefont{L.}~\bibnamefont{Anderson}}, \bibinfo
  {author} {\bibfnamefont{E.}~\bibnamefont{Aubourg}}, \bibinfo {author}
  {\bibfnamefont{S.}~\bibnamefont{Bailey}}, \bibinfo {author}
  {\bibfnamefont{D.}~\bibnamefont{Bizyaev}}, \bibinfo {author}
  {\bibfnamefont{M.}~\bibnamefont{Blanton}}, \emph{et~al.},\ }%
  \bibfield{journal}{%
  \Doi{10.1093/mnras/sts084}{\bibinfo {journal} {Mon.Not.Roy.Astron.Soc.}}\ }%
  \textbf{\bibinfo {volume} {428}},\ \bibinfo {pages} {1036} (\bibinfo {year}
  {2013}),\ \Eprint{http://arxiv.org/abs/1203.6594}{arXiv:1203.6594
  [astro-ph.CO]}%
  \bibAnnoteFile{NoStop}{Anderson:2012sa}%
\bibitem{Reid:2012sw}%
  \BibitemOpen
  \bibfield{author}{%
  \bibinfo {author} {\bibfnamefont{B.~A.}\ \bibnamefont{Reid}}, \bibinfo
  {author} {\bibfnamefont{L.}~\bibnamefont{Samushia}}, \bibinfo {author}
  {\bibfnamefont{M.}~\bibnamefont{White}}, \bibinfo {author}
  {\bibfnamefont{W.~J.}\ \bibnamefont{Percival}}, \bibinfo {author}
  {\bibfnamefont{M.}~\bibnamefont{Manera}}, \emph{et~al.}}%
   (\bibinfo {year} {2012}),\
  \Eprint{http://arxiv.org/abs/1203.6641}{arXiv:1203.6641 [astro-ph.CO]}%
  \bibAnnoteFile{NoStop}{Reid:2012sw}%
\bibitem{1974ApJ...187..425P}%
  \BibitemOpen
  \bibfield{author}{%
  \bibinfo {author} {\bibfnamefont{W.~H.}\ \bibnamefont{{Press}}}\ and\
  \bibinfo {author} {\bibfnamefont{P.}~\bibnamefont{{Schechter}}},\ }%
  \bibfield{journal}{%
  \Doi{10.1086/152650}{\bibinfo {journal} {\apj}}\ }%
  \textbf{\bibinfo {volume} {187}},\ \bibinfo {pages} {425} (\bibinfo {month}
  {Feb.}\ \bibinfo {year} {1974})%
  \bibAnnoteFile{NoStop}{1974ApJ...187..425P}%
\bibitem{Mo:1995cs}%
  \BibitemOpen
  \bibfield{author}{%
  \bibinfo {author} {\bibfnamefont{H.}~\bibnamefont{Mo}}\ and\ \bibinfo
  {author} {\bibfnamefont{S.~D.}\ \bibnamefont{White}},\ }%
  \bibfield{journal}{%
  \bibinfo {journal} {Mon.Not.Roy.Astron.Soc.}\ }%
  \textbf{\bibinfo {volume} {282}},\ \bibinfo {pages} {347} (\bibinfo {year}
  {1996}),\
  \Eprint{http://arxiv.org/abs/astro-ph/9512127}{arXiv:astro-ph/9512127
  [astro-ph]}%
  \bibAnnoteFile{NoStop}{Mo:1995cs}%
\bibitem{Fry:1992vr}%
  \BibitemOpen
  \bibfield{author}{%
  \bibinfo {author} {\bibfnamefont{J.~N.}\ \bibnamefont{Fry}}\ and\ \bibinfo
  {author} {\bibfnamefont{E.}~\bibnamefont{Gaztanaga}},\ }%
  \bibfield{journal}{%
  \Doi{10.1086/173015}{\bibinfo {journal} {Astrophys.J.}}\ }%
  \textbf{\bibinfo {volume} {413}},\ \bibinfo {pages} {447} (\bibinfo {year}
  {1993}),\
  \Eprint{http://arxiv.org/abs/astro-ph/9302009}{arXiv:astro-ph/9302009
  [astro-ph]}%
  \bibAnnoteFile{NoStop}{Fry:1992vr}%
\bibitem{Bullock:1999he}%
  \BibitemOpen
  \bibfield{author}{%
  \bibinfo {author} {\bibfnamefont{J.~S.}\ \bibnamefont{Bullock}}, \bibinfo
  {author} {\bibfnamefont{T.~S.}\ \bibnamefont{Kolatt}}, \bibinfo {author}
  {\bibfnamefont{Y.}~\bibnamefont{Sigad}}, \bibinfo {author}
  {\bibfnamefont{R.~S.}\ \bibnamefont{Somerville}}, \bibinfo {author}
  {\bibfnamefont{A.~V.}\ \bibnamefont{Kravtsov}}, \emph{et~al.},\ }%
  \bibfield{journal}{%
  \Doi{10.1046/j.1365-8711.2001.04068.x}{\bibinfo {journal}
  {Mon.Not.Roy.Astron.Soc.}}\ }%
  \textbf{\bibinfo {volume} {321}},\ \bibinfo {pages} {559} (\bibinfo {year}
  {2001}),\
  \Eprint{http://arxiv.org/abs/astro-ph/9908159}{arXiv:astro-ph/9908159
  [astro-ph]}%
  \bibAnnoteFile{NoStop}{Bullock:1999he}%
\bibitem{Neto:2007vq}%
  \BibitemOpen
  \bibfield{author}{%
  \bibinfo {author} {\bibfnamefont{A.~F.}\ \bibnamefont{Neto}}, \bibinfo
  {author} {\bibfnamefont{L.}~\bibnamefont{Gao}}, \bibinfo {author}
  {\bibfnamefont{P.}~\bibnamefont{Bett}}, \bibinfo {author}
  {\bibfnamefont{S.}~\bibnamefont{Cole}}, \bibinfo {author}
  {\bibfnamefont{J.~F.}\ \bibnamefont{Navarro}}, \emph{et~al.},\ }%
  \bibfield{journal}{%
  \Doi{10.1111/j.1365-2966.2007.12381.x}{\bibinfo {journal}
  {Mon.Not.Roy.Astron.Soc.}}\ }%
  \textbf{\bibinfo {volume} {381}},\ \bibinfo {pages} {1450} (\bibinfo {year}
  {2007}),\ \Eprint{http://arxiv.org/abs/0706.2919}{arXiv:0706.2919
  [astro-ph]}%
  \bibAnnoteFile{NoStop}{Neto:2007vq}%
\bibitem{Maccio':2008xb}%
  \BibitemOpen
  \bibfield{author}{%
  \bibinfo {author} {\bibfnamefont{A.~V.}\ \bibnamefont{Maccio'}}, \bibinfo
  {author} {\bibfnamefont{A.~A.}\ \bibnamefont{Dutton}},\ and\ \bibinfo
  {author} {\bibfnamefont{F.~C.~d.}\ \bibnamefont{Bosch}}}%
   (\bibinfo {year} {2008}),\
  \Eprint{http://arxiv.org/abs/0805.1926}{arXiv:0805.1926 [astro-ph]}%
  \bibAnnoteFile{NoStop}{Maccio':2008xb}%
\bibitem{Prada:2011jf}%
  \BibitemOpen
  \bibfield{author}{%
  \bibinfo {author} {\bibfnamefont{F.}~\bibnamefont{Prada}}, \bibinfo {author}
  {\bibfnamefont{A.~A.}\ \bibnamefont{Klypin}}, \bibinfo {author}
  {\bibfnamefont{A.~J.}\ \bibnamefont{Cuesta}}, \bibinfo {author}
  {\bibfnamefont{J.~E.}\ \bibnamefont{Betancort-Rijo}},\ and\ \bibinfo {author}
  {\bibfnamefont{J.}~\bibnamefont{Primack}}}%
   (\bibinfo {year} {2011}),\
  \Eprint{http://arxiv.org/abs/1104.5130}{arXiv:1104.5130 [astro-ph.CO]}%
  \bibAnnoteFile{NoStop}{Prada:2011jf}%
\bibitem{Li:2011vk}%
  \BibitemOpen
  \bibfield{author}{%
  \bibinfo {author} {\bibfnamefont{B.}~\bibnamefont{Li}}, \bibinfo {author}
  {\bibfnamefont{G.-B.}\ \bibnamefont{Zhao}}, \bibinfo {author}
  {\bibfnamefont{R.}~\bibnamefont{Teyssier}},\ and\ \bibinfo {author}
  {\bibfnamefont{K.}~\bibnamefont{Koyama}},\ }%
  \bibfield{journal}{%
  \Doi{10.1088/1475-7516/2012/01/051}{\bibinfo {journal} {JCAP}}\ }%
  \textbf{\bibinfo {volume} {1201}},\ \bibinfo {pages} {051} (\bibinfo {year}
  {2012}),\ \Eprint{http://arxiv.org/abs/1110.1379}{arXiv:1110.1379
  [astro-ph.CO]}%
  \bibAnnoteFile{NoStop}{Li:2011vk}%
\bibitem{Teyssier:2001cp}%
  \BibitemOpen
  \bibfield{author}{%
  \bibinfo {author} {\bibfnamefont{R.}~\bibnamefont{Teyssier}},\ }%
  \bibfield{journal}{%
  \Doi{10.1051/0004-6361:20011817}{\bibinfo {journal} {Astron.Astrophys.}}\ }%
  \textbf{\bibinfo {volume} {385}},\ \bibinfo {pages} {337} (\bibinfo {year}
  {2002}),\
  \Eprint{http://arxiv.org/abs/astro-ph/0111367}{arXiv:astro-ph/0111367
  [astro-ph]}%
  \bibAnnoteFile{NoStop}{Teyssier:2001cp}%
\bibitem{Li:2013nua}%
  \BibitemOpen
  \bibfield{author}{%
  \bibinfo {author} {\bibfnamefont{B.}~\bibnamefont{Li}}, \bibinfo {author}
  {\bibfnamefont{G.-B.}\ \bibnamefont{Zhao}},\ and\ \bibinfo {author}
  {\bibfnamefont{K.}~\bibnamefont{Koyama}},\ }%
  \bibfield{journal}{%
  \Doi{10.1088/1475-7516/2013/05/023}{\bibinfo {journal} {JCAP}}\ }%
  \textbf{\bibinfo {volume} {1305}},\ \bibinfo {pages} {023} (\bibinfo {year}
  {2013}),\ \Eprint{http://arxiv.org/abs/1303.0008}{arXiv:1303.0008
  [astro-ph.CO]}%
  \bibAnnoteFile{NoStop}{Li:2013nua}%
\bibitem{Behroozi:2011ju}%
  \BibitemOpen
  \bibfield{author}{%
  \bibinfo {author} {\bibfnamefont{P.~S.}\ \bibnamefont{Behroozi}}, \bibinfo
  {author} {\bibfnamefont{R.~H.}\ \bibnamefont{Wechsler}},\ and\ \bibinfo
  {author} {\bibfnamefont{H.-Y.}\ \bibnamefont{Wu}},\ }%
  \bibfield{journal}{%
  \Doi{10.1088/0004-637X/762/2/109}{\bibinfo {journal} {Astrophys.J.}}\ }%
  \textbf{\bibinfo {volume} {762}},\ \bibinfo {pages} {109} (\bibinfo {year}
  {2013}),\ \Eprint{http://arxiv.org/abs/1110.4372}{arXiv:1110.4372
  [astro-ph.CO]}%
  \bibAnnoteFile{NoStop}{Behroozi:2011ju}%
\bibitem{Wyman:2013jaa}%
  \BibitemOpen
  \bibfield{author}{%
  \bibinfo {author} {\bibfnamefont{M.}~\bibnamefont{Wyman}}, \bibinfo {author}
  {\bibfnamefont{E.}~\bibnamefont{Jennings}},\ and\ \bibinfo {author}
  {\bibfnamefont{M.}~\bibnamefont{Lima}},\ }%
  \bibfield{journal}{%
  \Doi{10.1103/PhysRevD.88.084029}{\bibinfo {journal} {Phys.Rev.}}\ }%
  \textbf{\bibinfo {volume} {D88}},\ \bibinfo {pages} {084029} (\bibinfo {year}
  {2013}),\ \Eprint{http://arxiv.org/abs/1303.6630}{arXiv:1303.6630
  [astro-ph.CO]}%
  \bibAnnoteFile{NoStop}{Wyman:2013jaa}%
\bibitem{Tinker:2008ff}%
  \BibitemOpen
  \bibfield{author}{%
  \bibinfo {author} {\bibfnamefont{J.~L.}\ \bibnamefont{Tinker}}, \bibinfo
  {author} {\bibfnamefont{A.~V.}\ \bibnamefont{Kravtsov}}, \bibinfo {author}
  {\bibfnamefont{A.}~\bibnamefont{Klypin}}, \bibinfo {author}
  {\bibfnamefont{K.}~\bibnamefont{Abazajian}}, \bibinfo {author}
  {\bibfnamefont{M.~S.}\ \bibnamefont{Warren}}, \emph{et~al.},\ }%
  \bibfield{journal}{%
  \Doi{10.1086/591439}{\bibinfo {journal} {Astrophys.J.}}\ }%
  \textbf{\bibinfo {volume} {688}},\ \bibinfo {pages} {709} (\bibinfo {year}
  {2008}),\ \Eprint{http://arxiv.org/abs/0803.2706}{arXiv:0803.2706
  [astro-ph]}%
  \bibAnnoteFile{NoStop}{Tinker:2008ff}%
\bibitem{Hellwing2009}%
  \BibitemOpen
  \bibfield{author}{%
  \bibinfo {author} {\bibfnamefont{W.~A.}\ \bibnamefont{{Hellwing}}}\ and\
  \bibinfo {author} {\bibfnamefont{R.}~\bibnamefont{{Juszkiewicz}}},\ }%
  \bibfield{journal}{%
  \Doi{10.1103/PhysRevD.80.083522}{\bibinfo {journal} {Phys.~Rev.~D}}\ }%
  \textbf{\bibinfo {volume} {80}},\ \bibinfo {eid} {083522} (\bibinfo {month}
  {Oct.}\ \bibinfo {year} {2009}),\
  \Eprint{http://arxiv.org/abs/0809.1976}{arXiv:0809.1976}%
  \bibAnnoteFile{NoStop}{Hellwing2009}%
\bibitem{Hellwing2010}%
  \BibitemOpen
  \bibfield{author}{%
  \bibinfo {author} {\bibfnamefont{W.~A.}\ \bibnamefont{{Hellwing}}}, \bibinfo
  {author} {\bibfnamefont{S.~R.}\ \bibnamefont{{Knollmann}}},\ and\ \bibinfo
  {author} {\bibfnamefont{A.}~\bibnamefont{{Knebe}}},\ }%
  \bibfield{journal}{%
  \Doi{10.1111/j.1745-3933.2010.00940.x}{\bibinfo {journal} {MNRAS}}\ }%
  \textbf{\bibinfo {volume} {408}},\ \bibinfo {pages} {L104} (\bibinfo {month}
  {Oct.}\ \bibinfo {year} {2010}),\
  \Eprint{http://arxiv.org/abs/1004.2929}{arXiv:1004.2929 [astro-ph.CO]}%
  \bibAnnoteFile{NoStop}{Hellwing2010}%
\bibitem{cv2011}%
  \BibitemOpen
  \bibfield{author}{%
  \bibinfo {author} {\bibfnamefont{M.~C.}\ \bibnamefont{{Cautun}}}\ and\
  \bibinfo {author} {\bibfnamefont{R.}~\bibnamefont{{van de Weygaert}}},\ }%
  \bibfield{journal}{%
  \bibinfo {journal} {ArXiv e-prints}}%
   (\bibinfo {month} {May}\ \bibinfo {year} {2011}),\
  \Eprint{http://arxiv.org/abs/1105.0370}{arXiv:1105.0370 [astro-ph.IM]}%
  \bibAnnoteFile{NoStop}{cv2011}%
\bibitem{sv2000}%
  \BibitemOpen
  \bibfield{author}{%
  \bibinfo {author} {\bibfnamefont{W.~E.}\ \bibnamefont{{Schaap}}}\ and\
  \bibinfo {author} {\bibfnamefont{R.}~\bibnamefont{{van de Weygaert}}},\ }%
  \bibfield{journal}{%
  \bibinfo {journal} {A\&A}\ }%
  \textbf{\bibinfo {volume} {363}},\ \bibinfo {pages} {L29} (\bibinfo {month}
  {Nov.}\ \bibinfo {year} {2000}),\
  \Eprint{http://arxiv.org/abs/arXiv:astro-ph/0011007}{arXiv:astro-ph/0011007}%
  \bibAnnoteFile{NoStop}{sv2000}%
\bibitem{Kravtsov:2003sg}%
  \BibitemOpen
  \bibfield{author}{%
  \bibinfo {author} {\bibfnamefont{A.~V.}\ \bibnamefont{Kravtsov}}, \bibinfo
  {author} {\bibfnamefont{A.~A.}\ \bibnamefont{Berlind}}, \bibinfo {author}
  {\bibfnamefont{R.~H.}\ \bibnamefont{Wechsler}}, \bibinfo {author}
  {\bibfnamefont{A.~A.}\ \bibnamefont{Klypin}}, \bibinfo {author}
  {\bibfnamefont{S.}~\bibnamefont{Gottloeber}}, \emph{et~al.},\ }%
  \bibfield{journal}{%
  \Doi{10.1086/420959}{\bibinfo {journal} {Astrophys.J.}}\ }%
  \textbf{\bibinfo {volume} {609}},\ \bibinfo {pages} {35} (\bibinfo {year}
  {2004}),\
  \Eprint{http://arxiv.org/abs/astro-ph/0308519}{arXiv:astro-ph/0308519
  [astro-ph]}%
  \bibAnnoteFile{NoStop}{Kravtsov:2003sg}%
\bibitem{Reid:2008zu}%
  \BibitemOpen
  \bibfield{author}{%
  \bibinfo {author} {\bibfnamefont{B.~A.}\ \bibnamefont{Reid}}, \bibinfo
  {author} {\bibfnamefont{D.~N.}\ \bibnamefont{Spergel}},\ and\ \bibinfo
  {author} {\bibfnamefont{P.}~\bibnamefont{Bode}},\ }%
  \bibfield{journal}{%
  \Doi{10.1088/0004-637X/702/1/249}{\bibinfo {journal} {Astrophys.J.}}\ }%
  \textbf{\bibinfo {volume} {702}},\ \bibinfo {pages} {249} (\bibinfo {year}
  {2009}),\ \Eprint{http://arxiv.org/abs/0811.1025}{arXiv:0811.1025
  [astro-ph]}%
  \bibAnnoteFile{NoStop}{Reid:2008zu}%
\bibitem{Wake:2008mf}%
  \BibitemOpen
  \bibfield{author}{%
  \bibinfo {author} {\bibfnamefont{D.~A.}\ \bibnamefont{Wake}}, \bibinfo
  {author} {\bibfnamefont{R.~K.}\ \bibnamefont{Sheth}}, \bibinfo {author}
  {\bibfnamefont{R.~C.}\ \bibnamefont{Nichol}}, \bibinfo {author}
  {\bibfnamefont{C.~M.}\ \bibnamefont{Baugh}}, \bibinfo {author}
  {\bibfnamefont{J.}~\bibnamefont{Bland-Hawthorn}}, \emph{et~al.},\ }%
  \bibfield{journal}{%
  \Doi{10.1111/j.1365-2966.2008.13333.x}{\bibinfo {journal}
  {Mon.Not.Roy.Astron.Soc.}}\ }%
  \textbf{\bibinfo {volume} {387}},\ \bibinfo {pages} {1045} (\bibinfo {year}
  {2008}),\ \Eprint{http://arxiv.org/abs/0802.4288}{arXiv:0802.4288
  [astro-ph]}%
  \bibAnnoteFile{NoStop}{Wake:2008mf}%
\bibitem{Zheng:2008np}%
  \BibitemOpen
  \bibfield{author}{%
  \bibinfo {author} {\bibfnamefont{Z.}~\bibnamefont{Zheng}}, \bibinfo {author}
  {\bibfnamefont{I.}~\bibnamefont{Zehavi}}, \bibinfo {author}
  {\bibfnamefont{D.~J.}\ \bibnamefont{Eisenstein}}, \bibinfo {author}
  {\bibfnamefont{D.~H.}\ \bibnamefont{Weinberg}},\ and\ \bibinfo {author}
  {\bibfnamefont{Y.}~\bibnamefont{Jing}},\ }%
  \bibfield{journal}{%
  \Doi{10.1088/0004-637X/707/1/554}{\bibinfo {journal} {Astrophys.J.}}\ }%
  \textbf{\bibinfo {volume} {707}},\ \bibinfo {pages} {554} (\bibinfo {year}
  {2009}),\ \Eprint{http://arxiv.org/abs/0809.1868}{arXiv:0809.1868
  [astro-ph]}%
  \bibAnnoteFile{NoStop}{Zheng:2008np}%
\bibitem{Sawangwit:2009bg}%
  \BibitemOpen
  \bibfield{author}{%
  \bibinfo {author} {\bibfnamefont{U.}~\bibnamefont{Sawangwit}}, \bibinfo
  {author} {\bibfnamefont{T.}~\bibnamefont{Shanks}}, \bibinfo {author}
  {\bibfnamefont{F.}~\bibnamefont{Abdalla}}, \bibinfo {author}
  {\bibfnamefont{R.}~\bibnamefont{Cannon}}, \bibinfo {author}
  {\bibfnamefont{S.}~\bibnamefont{Croom}}, \emph{et~al.},\ }%
  \bibfield{journal}{%
  \Doi{10.1111/j.1365-2966.2011.19251.x}{\bibinfo {journal}
  {Mon.Not.Roy.Astron.Soc.}}\ }%
  \textbf{\bibinfo {volume} {416}},\ \bibinfo {pages} {3033} (\bibinfo {year}
  {2011}),\ \Eprint{http://arxiv.org/abs/0912.0511}{arXiv:0912.0511
  [astro-ph.CO]}%
  \bibAnnoteFile{NoStop}{Sawangwit:2009bg}%
\bibitem{Reid11112011}%
  \BibitemOpen
  \bibfield{author}{%
  \bibinfo {author} {\bibfnamefont{B.~A.}\ \bibnamefont{Reid}} \emph{et~al.},\
  }%
  \bibfield{journal}{%
  \Doi{10.1111/j.1365-2966.2011.18943.x}{\bibinfo {journal} {Monthly Notices of
  the Royal Astronomical Society}}\ }%
  \textbf{\bibinfo {volume} {417}},\ \bibinfo {pages} {3103} (\bibinfo {year}
  {2011})%
  \bibAnnoteFile{NoStop}{Reid11112011}%
\bibitem{Baugh:2006pf}%
  \BibitemOpen
  \bibfield{author}{%
  \bibinfo {author} {\bibfnamefont{C.~M.}\ \bibnamefont{Baugh}},\ }%
  \bibfield{journal}{%
  \Doi{10.1088/0034-4885/69/12/R02}{\bibinfo {journal} {Rept.Prog.Phys.}}\ }%
  \textbf{\bibinfo {volume} {69}},\ \bibinfo {pages} {3101} (\bibinfo {year}
  {2006}),\
  \Eprint{http://arxiv.org/abs/astro-ph/0610031}{arXiv:astro-ph/0610031
  [astro-ph]}%
  \bibAnnoteFile{NoStop}{Baugh:2006pf}%
\bibitem{Almeida:2007ef}%
  \BibitemOpen
  \bibfield{author}{%
  \bibinfo {author} {\bibfnamefont{C.}~\bibnamefont{Almeida}}, \bibinfo
  {author} {\bibfnamefont{C.}~\bibnamefont{Baugh}}, \bibinfo {author}
  {\bibfnamefont{D.}~\bibnamefont{Wake}}, \bibinfo {author}
  {\bibfnamefont{C.}~\bibnamefont{Lacey}}, \bibinfo {author}
  {\bibfnamefont{A.}~\bibnamefont{Benson}}, \emph{et~al.},\ }%
  \bibfield{journal}{%
  \bibinfo {journal} {: Mon.Not.Roy.Astron.Soc.}}%
   (\bibinfo {year} {2007}),\
  \Eprint{http://arxiv.org/abs/0710.3557}{arXiv:0710.3557 [astro-ph]}%
  \bibAnnoteFile{NoStop}{Almeida:2007ef}%
\bibitem{Banerji:2009pv}%
  \BibitemOpen
  \bibfield{author}{%
  \bibinfo {author} {\bibfnamefont{M.}~\bibnamefont{Banerji}}, \bibinfo
  {author} {\bibfnamefont{I.}~\bibnamefont{Ferreras}}, \bibinfo {author}
  {\bibfnamefont{F.~B.}\ \bibnamefont{Abdalla}}, \bibinfo {author}
  {\bibfnamefont{P.}~\bibnamefont{Hewett}},\ and\ \bibinfo {author}
  {\bibfnamefont{O.}~\bibnamefont{Lahav}},\ }%
  \bibfield{journal}{%
  \Doi{10.1111/j.1365-2966.2009.16060.x}{\bibinfo {journal}
  {Mon.Not.Roy.Astron.Soc.}}\ }%
  \textbf{\bibinfo {volume} {402}},\ \bibinfo {pages} {2264} (\bibinfo {year}
  {2010}),\ \Eprint{http://arxiv.org/abs/0910.5372}{arXiv:0910.5372
  [astro-ph.CO]}%
  \bibAnnoteFile{NoStop}{Banerji:2009pv}%
\bibitem{Maccio:2008xb}%
  \BibitemOpen
  \bibfield{author}{%
  \bibinfo {author} {\bibfnamefont{A.~V.}\ \bibnamefont{Maccio'}}, \bibinfo
  {author} {\bibfnamefont{A.~A.}\ \bibnamefont{Dutton}},\ and\ \bibinfo
  {author} {\bibfnamefont{F.~C.~d.}\ \bibnamefont{Bosch}}}%
   (\bibinfo {year} {2008}),\
  \Eprint{http://arxiv.org/abs/0805.1926}{arXiv:0805.1926 [astro-ph]}%
  \bibAnnoteFile{NoStop}{Maccio:2008xb}%
\bibitem{Komatsu:2008hk}%
  \BibitemOpen
  \bibfield{author}{%
  \bibinfo {author} {\bibfnamefont{E.}~\bibnamefont{Komatsu}} \emph{et~al.}
  (\bibinfo {collaboration} {WMAP Collaboration}),\ }%
  \bibfield{journal}{%
  \Doi{10.1088/0067-0049/180/2/330}{\bibinfo {journal} {Astrophys.J.Suppl.}}\
  }%
  \textbf{\bibinfo {volume} {180}},\ \bibinfo {pages} {330} (\bibinfo {year}
  {2009}),\ \Eprint{http://arxiv.org/abs/0803.0547}{arXiv:0803.0547
  [astro-ph]}%
  \bibAnnoteFile{NoStop}{Komatsu:2008hk}%
\bibitem{MunozCuartas:2010ig}%
  \BibitemOpen
  \bibfield{author}{%
  \bibinfo {author} {\bibfnamefont{J.}~\bibnamefont{Munoz-Cuartas}}, \bibinfo
  {author} {\bibfnamefont{A.}~\bibnamefont{Maccio}}, \bibinfo {author}
  {\bibfnamefont{S.}~\bibnamefont{Gottlober}},\ and\ \bibinfo {author}
  {\bibfnamefont{A.}~\bibnamefont{Dutton}}}%
   (\bibinfo {year} {2010}),\
  \Eprint{http://arxiv.org/abs/1007.0438}{arXiv:1007.0438 [astro-ph.CO]}%
  \bibAnnoteFile{NoStop}{MunozCuartas:2010ig}%
\bibitem{Ludlow:2013vxa}%
  \BibitemOpen
  \bibfield{author}{%
  \bibinfo {author} {\bibfnamefont{A.~D.}\ \bibnamefont{Ludlow}}, \bibinfo
  {author} {\bibfnamefont{J.~F.}\ \bibnamefont{Navarro}}, \bibinfo {author}
  {\bibfnamefont{R.~E.}\ \bibnamefont{Angulo}}, \bibinfo {author}
  {\bibfnamefont{M.}~\bibnamefont{Boylan-Kolchin}}, \bibinfo {author}
  {\bibfnamefont{V.}~\bibnamefont{Springel}}, \emph{et~al.}}%
   (\bibinfo {year} {2013}),\
  \Eprint{http://arxiv.org/abs/1312.0945}{arXiv:1312.0945 [astro-ph.CO]}%
  \bibAnnoteFile{NoStop}{Ludlow:2013vxa}%
\bibitem{Hellwing2010b}%
  \BibitemOpen
  \bibfield{author}{%
  \bibinfo {author} {\bibfnamefont{W.~A.}\ \bibnamefont{{Hellwing}}},\ }%
  \bibfield{journal}{%
  \Doi{10.1002/andp.201010445}{\bibinfo {journal} {Annalen der Physik}}\ }%
  \textbf{\bibinfo {volume} {522}},\ \bibinfo {pages} {351} (\bibinfo {month}
  {Mar.}\ \bibinfo {year} {2010}),\
  \Eprint{http://arxiv.org/abs/0911.0573}{arXiv:0911.0573 [astro-ph.CO]}%
  \bibAnnoteFile{NoStop}{Hellwing2010b}%
\bibitem{Hellwing:2008qf}%
  \BibitemOpen
  \bibfield{author}{%
  \bibinfo {author} {\bibfnamefont{W.~A.}\ \bibnamefont{Hellwing}}\ and\
  \bibinfo {author} {\bibfnamefont{R.}~\bibnamefont{Juszkiewicz}},\ }%
  \bibfield{journal}{%
  \Doi{10.1103/PhysRevD.80.083522}{\bibinfo {journal} {Phys.Rev.}}\ }%
  \textbf{\bibinfo {volume} {D80}},\ \bibinfo {pages} {083522} (\bibinfo {year}
  {2009}),\ \Eprint{http://arxiv.org/abs/0809.1976}{arXiv:0809.1976
  [astro-ph]}%
  \bibAnnoteFile{NoStop}{Hellwing:2008qf}%
\bibitem{Sakstein:2013pda}%
  \BibitemOpen
  \bibfield{author}{%
  \bibinfo {author} {\bibfnamefont{J.}~\bibnamefont{Sakstein}}}%
   (\bibinfo {year} {2013}),\
  \Eprint{http://arxiv.org/abs/1309.0495}{arXiv:1309.0495 [astro-ph.CO]}%
  \bibAnnoteFile{NoStop}{Sakstein:2013pda}%
\bibitem{Khoury:2003rn}%
  \BibitemOpen
  \bibfield{author}{%
  \bibinfo {author} {\bibfnamefont{J.}~\bibnamefont{Khoury}}\ and\ \bibinfo
  {author} {\bibfnamefont{A.}~\bibnamefont{Weltman}},\ }%
  \bibfield{journal}{%
  \Doi{10.1103/PhysRevD.69.044026}{\bibinfo {journal} {Phys.Rev.}}\ }%
  \textbf{\bibinfo {volume} {D69}},\ \bibinfo {pages} {044026} (\bibinfo {year}
  {2004}),\
  \Eprint{http://arxiv.org/abs/astro-ph/0309411}{arXiv:astro-ph/0309411
  [astro-ph]}%
  \bibAnnoteFile{NoStop}{Khoury:2003rn}%
\bibitem{Mota:2006fz}%
  \BibitemOpen
  \bibfield{author}{%
  \bibinfo {author} {\bibfnamefont{D.~F.}\ \bibnamefont{Mota}}\ and\ \bibinfo
  {author} {\bibfnamefont{D.~J.}\ \bibnamefont{Shaw}},\ }%
  \bibfield{journal}{%
  \Doi{10.1103/PhysRevD.75.063501}{\bibinfo {journal} {Phys.Rev.}}\ }%
  \textbf{\bibinfo {volume} {D75}},\ \bibinfo {pages} {063501} (\bibinfo {year}
  {2007}),\ \Eprint{http://arxiv.org/abs/hep-ph/0608078}{arXiv:hep-ph/0608078
  [hep-ph]}%
  \bibAnnoteFile{NoStop}{Mota:2006fz}%
\bibitem{Colombi:2008dw}%
  \BibitemOpen
  \bibfield{author}{%
  \bibinfo {author} {\bibfnamefont{S.}~\bibnamefont{Colombi}}, \bibinfo
  {author} {\bibfnamefont{A.~H.}\ \bibnamefont{Jaffe}}, \bibinfo {author}
  {\bibfnamefont{D.}~\bibnamefont{Novikov}},\ and\ \bibinfo {author}
  {\bibfnamefont{C.}~\bibnamefont{Pichon}}}%
   (\bibinfo {year} {2008}),\
  \Eprint{http://arxiv.org/abs/0811.0313}{arXiv:0811.0313 [astro-ph]}%
  \bibAnnoteFile{NoStop}{Colombi:2008dw}%
\bibitem{Mo:1996cn}%
  \BibitemOpen
  \bibfield{author}{%
  \bibinfo {author} {\bibfnamefont{H.}~\bibnamefont{Mo}}, \bibinfo {author}
  {\bibfnamefont{Y.}~\bibnamefont{Jing}},\ and\ \bibinfo {author}
  {\bibfnamefont{S.}~\bibnamefont{White}}}%
   (\bibinfo {year} {1996}),\
  \Eprint{http://arxiv.org/abs/astro-ph/9603039}{arXiv:astro-ph/9603039
  [astro-ph]}%
  \bibAnnoteFile{NoStop}{Mo:1996cn}%
\bibitem{Scoccimarro:2000gm}%
  \BibitemOpen
  \bibfield{author}{%
  \bibinfo {author} {\bibfnamefont{R.}~\bibnamefont{Scoccimarro}}, \bibinfo
  {author} {\bibfnamefont{R.~K.}\ \bibnamefont{Sheth}}, \bibinfo {author}
  {\bibfnamefont{L.}~\bibnamefont{Hui}},\ and\ \bibinfo {author}
  {\bibfnamefont{B.}~\bibnamefont{Jain}},\ }%
  \bibfield{journal}{%
  \Doi{10.1086/318261}{\bibinfo {journal} {Astrophys.J.}}\ }%
  \textbf{\bibinfo {volume} {546}},\ \bibinfo {pages} {20} (\bibinfo {year}
  {2001}),\
  \Eprint{http://arxiv.org/abs/astro-ph/0006319}{arXiv:astro-ph/0006319
  [astro-ph]}%
  \bibAnnoteFile{NoStop}{Scoccimarro:2000gm}%
\bibitem{Lombriser:2013eza}%
  \BibitemOpen
  \bibfield{author}{%
  \bibinfo {author} {\bibfnamefont{L.}~\bibnamefont{Lombriser}}, \bibinfo
  {author} {\bibfnamefont{K.}~\bibnamefont{Koyama}},\ and\ \bibinfo {author}
  {\bibfnamefont{B.}~\bibnamefont{Li}}}%
   (\bibinfo {year} {2013}),\
  \Eprint{http://arxiv.org/abs/1312.1292}{arXiv:1312.1292 [astro-ph.CO]}%
  \bibAnnoteFile{NoStop}{Lombriser:2013eza}%
\bibitem{Smith:2002dz}%
  \BibitemOpen
  \bibfield{author}{%
  \bibinfo {author} {\bibfnamefont{R.}~\bibnamefont{Smith}} \emph{et~al.}
  (\bibinfo {collaboration} {Virgo Consortium}),\ }%
  \bibfield{journal}{%
  \Doi{10.1046/j.1365-8711.2003.06503.x}{\bibinfo {journal}
  {Mon.Not.Roy.Astron.Soc.}}\ }%
  \textbf{\bibinfo {volume} {341}},\ \bibinfo {pages} {1311} (\bibinfo {year}
  {2003}),\
  \Eprint{http://arxiv.org/abs/astro-ph/0207664}{arXiv:astro-ph/0207664
  [astro-ph]}%
  \bibAnnoteFile{NoStop}{Smith:2002dz}%
\bibitem{Takahashi:2012em}%
  \BibitemOpen
  \bibfield{author}{%
  \bibinfo {author} {\bibfnamefont{R.}~\bibnamefont{Takahashi}}, \bibinfo
  {author} {\bibfnamefont{M.}~\bibnamefont{Sato}}, \bibinfo {author}
  {\bibfnamefont{T.}~\bibnamefont{Nishimichi}}, \bibinfo {author}
  {\bibfnamefont{A.}~\bibnamefont{Taruya}},\ and\ \bibinfo {author}
  {\bibfnamefont{M.}~\bibnamefont{Oguri}},\ }%
  \bibfield{journal}{%
  \Doi{10.1088/0004-637X/761/2/152}{\bibinfo {journal} {Astrophys.J.}}\ }%
  \textbf{\bibinfo {volume} {761}},\ \bibinfo {pages} {152} (\bibinfo {year}
  {2012}),\ \Eprint{http://arxiv.org/abs/1208.2701}{arXiv:1208.2701
  [astro-ph.CO]}%
  \bibAnnoteFile{NoStop}{Takahashi:2012em}%
\bibitem{Zhao:2013dza}%
  \BibitemOpen
  \bibfield{author}{%
  \bibinfo {author} {\bibfnamefont{G.-B.}\ \bibnamefont{Zhao}}}%
   (\bibinfo {year} {2013}),\
  \Eprint{http://arxiv.org/abs/1312.1291}{arXiv:1312.1291 [astro-ph.CO]}%
  \bibAnnoteFile{NoStop}{Zhao:2013dza}%
\bibitem{Schmidt:2009yj}%
  \BibitemOpen
  \bibfield{author}{%
  \bibinfo {author} {\bibfnamefont{F.}~\bibnamefont{Schmidt}}, \bibinfo
  {author} {\bibfnamefont{W.}~\bibnamefont{Hu}},\ and\ \bibinfo {author}
  {\bibfnamefont{M.}~\bibnamefont{Lima}},\ }%
  \bibfield{journal}{%
  \Doi{10.1103/PhysRevD.81.063005}{\bibinfo {journal} {Phys.Rev.}}\ }%
  \textbf{\bibinfo {volume} {D81}},\ \bibinfo {pages} {063005} (\bibinfo {year}
  {2010}),\ \Eprint{http://arxiv.org/abs/0911.5178}{arXiv:0911.5178
  [astro-ph.CO]}%
  \bibAnnoteFile{NoStop}{Schmidt:2009yj}%
\bibitem{Schmidt:2008tn}%
  \BibitemOpen
  \bibfield{author}{%
  \bibinfo {author} {\bibfnamefont{F.}~\bibnamefont{Schmidt}}, \bibinfo
  {author} {\bibfnamefont{M.~V.}\ \bibnamefont{Lima}}, \bibinfo {author}
  {\bibfnamefont{H.}~\bibnamefont{Oyaizu}},\ and\ \bibinfo {author}
  {\bibfnamefont{W.}~\bibnamefont{Hu}},\ }%
  \bibfield{journal}{%
  \Doi{10.1103/PhysRevD.79.083518}{\bibinfo {journal} {Phys.Rev.}}\ }%
  \textbf{\bibinfo {volume} {D79}},\ \bibinfo {pages} {083518} (\bibinfo {year}
  {2009}),\ \Eprint{http://arxiv.org/abs/0812.0545}{arXiv:0812.0545
  [astro-ph]}%
  \bibAnnoteFile{NoStop}{Schmidt:2008tn}%
\bibitem{Gaztanaga:2000vw}%
  \BibitemOpen
  \bibfield{author}{%
  \bibinfo {author} {\bibfnamefont{E.}~\bibnamefont{Gaztanaga}}\ and\ \bibinfo
  {author} {\bibfnamefont{J.~A.}\ \bibnamefont{Lobo}},\ }%
  \bibfield{journal}{%
  \Doi{10.1086/318684}{\bibinfo {journal} {Astrophys.J.}}\ }%
  \textbf{\bibinfo {volume} {548}},\ \bibinfo {pages} {47} (\bibinfo {year}
  {2001}),\
  \Eprint{http://arxiv.org/abs/astro-ph/0003129}{arXiv:astro-ph/0003129
  [astro-ph]}%
  \bibAnnoteFile{NoStop}{Gaztanaga:2000vw}%
\bibitem{Schaefer:2007nf}%
  \BibitemOpen
  \bibfield{author}{%
  \bibinfo {author} {\bibfnamefont{B.~M.}\ \bibnamefont{Schaefer}}\ and\
  \bibinfo {author} {\bibfnamefont{K.}~\bibnamefont{Koyama}},\ }%
  \bibfield{journal}{%
  \Doi{10.1111/j.1365-2966.2008.12841.x}{\bibinfo {journal}
  {Mon.Not.Roy.Astron.Soc.}}\ }%
  \textbf{\bibinfo {volume} {385}},\ \bibinfo {pages} {411} (\bibinfo {year}
  {2008}),\ \Eprint{http://arxiv.org/abs/0711.3129}{arXiv:0711.3129
  [astro-ph]}%
  \bibAnnoteFile{NoStop}{Schaefer:2007nf}%
\bibitem{Martino:2008ae}%
  \BibitemOpen
  \bibfield{author}{%
  \bibinfo {author} {\bibfnamefont{M.~C.}\ \bibnamefont{Martino}}, \bibinfo
  {author} {\bibfnamefont{H.~F.}\ \bibnamefont{Stabenau}},\ and\ \bibinfo
  {author} {\bibfnamefont{R.~K.}\ \bibnamefont{Sheth}},\ }%
  \bibfield{journal}{%
  \Doi{10.1103/PhysRevD.79.084013}{\bibinfo {journal} {Phys.Rev.}}\ }%
  \textbf{\bibinfo {volume} {D79}},\ \bibinfo {pages} {084013} (\bibinfo {year}
  {2009}),\ \Eprint{http://arxiv.org/abs/0812.0200}{arXiv:0812.0200
  [astro-ph]}%
  \bibAnnoteFile{NoStop}{Martino:2008ae}%
\bibitem{Li:2011qda}%
  \BibitemOpen
  \bibfield{author}{%
  \bibinfo {author} {\bibfnamefont{B.}~\bibnamefont{Li}}\ and\ \bibinfo
  {author} {\bibfnamefont{G.}~\bibnamefont{Efstathiou}},\ }%
  \bibfield{journal}{%
  \Doi{10.1111/j.1365-2966.2011.20404.x}{\bibinfo {journal}
  {Mon.Not.Roy.Astron.Soc.}}\ }%
  \textbf{\bibinfo {volume} {421}},\ \bibinfo {pages} {1431} (\bibinfo {year}
  {2012}),\ \Eprint{http://arxiv.org/abs/1110.6440}{arXiv:1110.6440
  [astro-ph.CO]}%
  \bibAnnoteFile{NoStop}{Li:2011qda}%
\bibitem{Borisov:2011fu}%
  \BibitemOpen
  \bibfield{author}{%
  \bibinfo {author} {\bibfnamefont{A.}~\bibnamefont{Borisov}}, \bibinfo
  {author} {\bibfnamefont{B.}~\bibnamefont{Jain}},\ and\ \bibinfo {author}
  {\bibfnamefont{P.}~\bibnamefont{Zhang}},\ }%
  \bibfield{journal}{%
  \Doi{10.1103/PhysRevD.85.063518}{\bibinfo {journal} {Phys.Rev.}}\ }%
  \textbf{\bibinfo {volume} {D85}},\ \bibinfo {pages} {063518} (\bibinfo {year}
  {2012}),\ \Eprint{http://arxiv.org/abs/1102.4839}{arXiv:1102.4839
  [astro-ph.CO]}%
  \bibAnnoteFile{NoStop}{Borisov:2011fu}%
\bibitem{Lombriser:2013wta}%
  \BibitemOpen
  \bibfield{author}{%
  \bibinfo {author} {\bibfnamefont{L.}~\bibnamefont{Lombriser}}, \bibinfo
  {author} {\bibfnamefont{B.}~\bibnamefont{Li}}, \bibinfo {author}
  {\bibfnamefont{K.}~\bibnamefont{Koyama}},\ and\ \bibinfo {author}
  {\bibfnamefont{G.-B.}\ \bibnamefont{Zhao}},\ }%
  \bibfield{journal}{%
  \bibinfo {journal} {Phys. Rev. D 87,}\ }%
  \textbf{\bibinfo {volume} {123511}} (\bibinfo {year} {2013}),\ \doi{\bibinfo
  {doi} {10.1103/PhysRevD.87.123511}},\
  \Eprint{http://arxiv.org/abs/1304.6395}{arXiv:1304.6395 [astro-ph.CO]}%
  \bibAnnoteFile{NoStop}{Lombriser:2013wta}%
\bibitem{Kopp:2013lea}%
  \BibitemOpen
  \bibfield{author}{%
  \bibinfo {author} {\bibfnamefont{M.}~\bibnamefont{Kopp}}, \bibinfo {author}
  {\bibfnamefont{S.~A.}\ \bibnamefont{Appleby}}, \bibinfo {author}
  {\bibfnamefont{I.}~\bibnamefont{Achitouv}},\ and\ \bibinfo {author}
  {\bibfnamefont{J.}~\bibnamefont{Weller}}}%
   (\bibinfo {year} {2013}),\
  \Eprint{http://arxiv.org/abs/1306.3233}{arXiv:1306.3233 [astro-ph.CO]}%
  \bibAnnoteFile{NoStop}{Kopp:2013lea}%
\bibitem{Taddei:2013bsk}%
  \BibitemOpen
  \bibfield{author}{%
  \bibinfo {author} {\bibfnamefont{L.}~\bibnamefont{Taddei}}, \bibinfo {author}
  {\bibfnamefont{R.}~\bibnamefont{Catena}},\ and\ \bibinfo {author}
  {\bibfnamefont{M.}~\bibnamefont{Pietroni}}}%
   (\bibinfo {year} {2013}),\
  \Eprint{http://arxiv.org/abs/1310.6175}{arXiv:1310.6175 [astro-ph.CO]}%
  \bibAnnoteFile{NoStop}{Taddei:2013bsk}%
\end{thebibliography}%

\end{document}